\documentclass[superscriptaddress,aps,prx,nofootinbib,notitlepage,10pt,longbibliography,reprint]{revtex4-2}  % two-column
\usepackage{amsmath}
\usepackage{amssymb}
\usepackage{physics}
\usepackage{graphicx}
\usepackage{color}
\usepackage{float}
\usepackage{booktabs}
\usepackage{multirow}
\usepackage{bbm}
\usepackage{listings}
\usepackage[table,xcdraw]{xcolor}
\usepackage[
    colorlinks=true,
    citecolor=blue,
    linkcolor=blue,
	filecolor=blue,
	urlcolor=blue,
]{hyperref}
\newcommand{\gv}[1]{\ensuremath{\mbox{\boldmath$ #1 $}}}
\newcommand{\avg}[1]{\left< #1 \right>}

\begin{document}

\title{Modeling quantum information scrambling in realistic protein NMR experiments}

\author{Alec F. White}
\affiliation{%
Google Quantum AI, San Francisco, CA, USA}
\author{Joseph Haslehurst}
\affiliation{School of Chemistry and Chemical Engineering, University of Southampton, Southampton, UK}
\author{Daniel R. Moberg}
\affiliation{%
Quantum Simulation Technologies Inc., Boston, MA, USA}
\author{Justin Provazza}
\affiliation{%
Quantum Simulation Technologies Inc., Boston, MA, USA}
\author{Nicholas C. Rubin}
\affiliation{%
Google Quantum AI, San Francisco, CA, USA}
\author{Thomas E. O'Brien}
\affiliation{%
Google Quantum AI, San Francisco, CA, USA}
\author{Christian Bengs}
\affiliation{School of Chemistry and Chemical Engineering, University of Southampton, Southampton, UK}

\begin{abstract}
Out-of-time-order correlators (OTOCs) are important probes of operator spreading and information scrambling, and have recently been proposed as a means of extracting molecular structural information through nuclear magnetic resonance (NMR). Realizing this potential requires accurate modeling of OTOCs in complex molecular environments within experimentally relevant settings. Here, we develop an end-to-end computational pipeline connecting molecular structures to OTOC dynamics in protein systems. Our framework incorporates molecular motion through molecular dynamics simulations, orientational averaging appropriate for powder samples, and spin-network truncation, using an experimentally realizable double-quantum effective Hamiltonian. We apply this framework to a dataset of over 350 carbon--carbon spin pairs across 10 protein systems, spanning interspin distances of $5$--$17$~\AA. We find that OTOC arrival times scale approximately linearly with interspin distance, revealing system-dependent operator-spreading velocities that enable correlations to propagate over extended distances within milliseconds. Our sensitivity analysis further shows that individual OTOCs predominantly constrain a single structural parameter, primarily associated with interspin distance. Together, these findings highlight the potential of OTOCs to probe spatially extended spin correlations on timescales compatible with solid-state NMR experiments.
\end{abstract}
\maketitle

\tableofcontents

%%%%%%%%%%%%%%%%%%%%%%%%%%%%%%%%%%%%%%%%%
\section{Introduction}\label{sec:intro}
%%%%%%%%%%%%%%%%%%%%%%%%%%%%%%%%%%%%%%%%%
Out-of-time-order correlators (OTOCs) have emerged as important probes of operator spreading, information scrambling, and quantum chaos in many-body systems~\cite{zhou2023operator,maldacenaBoundChaos2016,shenkerBlackHolesButterfly2014,hosurChaosQuantumChannels2016,swingleMeasuringScramblingQuantum2016a,swingleUnscramblingPhysicsOutoftimeorder2018b,togaInformationPropagationRydberg2026,liErrorResilientReversalQuantum2026}. More recently, OTOCs have been proposed as a means of characterizing interactions in strongly coupled quantum systems, offering particular advantages for probing weak couplings and spatially extended correlations~\cite{obrien2022quantum,schuster2023learning,google25constructive}. Solid-state nuclear magnetic resonance (NMR) provides a natural experimental setting for this approach, building on related studies of operator spreading using multiple-quantum coherence spectroscopy~\cite{cho_decay_2006,alvarez2015localization,dominguez21dynamicz} and Loschmidt echo experiments~\cite{Levstein1998,PASTAWSKI2000166}. In this setting, the rapid spreading of operators through nuclear spin networks may enable access to longer-range structural information on shorter timescales~\cite{zhang2025quantum}. Such long-range information is particularly valuable for constraining the overall molecular conformation, complementing the local structural information provided by short-range measurements. Realizing this potential, however, requires establishing a quantitative connection between molecular structure and OTOC dynamics under experimentally relevant conditions.

For generic interacting spin systems, no efficient classical method for simulating OTOCs is known, motivating their simulation on quantum computers, as envisioned in related NMR proposals~\cite{sels2020quantum,seetharam2023digital,burov2025largecircuitexecutionnmr,walch2026effectsshotnoisequantum}. However, simulating an OTOC for a given Hamiltonian addresses only part of the computational challenge. In solid-state NMR, local molecular motion modifies the dipolar interactions between nuclear spins, while powder samples require averaging over molecular orientations~\cite{SCHANDA20161,Eden_PA_2003}. Furthermore, large proteins and protein assemblies can contain tens of thousands of proton spins, making an exact quantum-mechanical description computationally intractable and motivating the use of reduced spin systems or approximate treatments of the spin dynamics~\cite{KUPROV2007241,KUPROV200845,Dumez_spintruncation_2010,EDWARDS2014107,Grasser_spinDMFT_2026}. These challenges parallel those encountered in quantum electronic structure calculations, where substantial classical preprocessing is required before the quantum computation~\cite{reiher2017elucidating,lee2021even,goings2022reliably}. Developing a comparable pipeline for NMR OTOC simulation is therefore a crucial step toward connecting molecular structures to experimentally accessible spin dynamics, providing the forward model needed to address the inverse problem of structural determination.

Here, we develop such a framework for modeling OTOCs in protein systems under conditions relevant to solid-state NMR experiments. Starting from atomic structures, we incorporate fast molecular motion through short molecular dynamics trajectories, extract effective nuclear-spin couplings, and perform the orientational averaging appropriate for powder samples. Motivated by recent work\cite{zhang2025quantum}, we focus on an experimentally realizable double-quantum effective Hamiltonian and develop methods to reduce the spin simulation problem by truncating away unnecessary spins. Using this framework, we construct a dataset of more than 350 pairs of measurement and butterfly spins, spanning distances of $5$--$17$~\AA~across 10 protein systems. We find that the OTOC arrival time scales approximately linearly with interspin distance, allowing us to estimate a system-dependent operator-spreading velocity. To investigate the structural information encoded in these dynamics, we analyze the sensitivity of OTOCs to small geometric perturbations in a simplified proxy system, quantified using a Fisher-information-based metric. Our results suggest that a single OTOC constrains one structural parameter, predominantly associated with the interspin distance. We further benchmark several approximate classical simulation methods, including Pauli propagation, truncated Liouvillian simulation, matrix product operators, and the discrete truncated Wigner approximation. We find that none of these methods achieves sufficient accuracy at a computational cost low enough to reliably interpret experimental OTOCs in the regimes studied. Together, these results establish a computational framework connecting molecular structure to OTOCs under realistic NMR conditions, providing a foundation for future structure reconstruction efforts while identifying the computational and information-theoretic challenges that remain.

%%%%%%%%%%%%%%%%%%%%%%%%%%%%%%%%%%%%%%%%%
\section{Out-of-Time-Order Correlators in Protein NMR}\label{sec:theory}
%%%%%%%%%%%%%%%%%%%%%%%%%%%%%%%%%%%%%%%%%

%%%%%%%%%%%%%%%%%%%%%%%%%%%%%%%%%%%%%%%%%%
%\subsection{The out-of-time-order correlator}
%%%%%%%%%%%%%%%%%%%%%%%%%%%%%%%%%%%%%%%%%%
The infinite temperature OTOCs that we will consider here have the form
\begin{equation}\label{eq:OTOC}
    \mathrm{OTOC} = \frac{1}{2^n} \mathrm{Tr}\left[B^{\dagger}(t)M^{\dagger} B(t) M  \right],
\end{equation}
or some cyclic permutation under which the trace is invariant. Here and elsewhere we use $n$ to indicate the total number of spins in the problem. To connect this definition to the target physical systems, we envision two labeled ${}^{13}$C atoms, $b$ and $m$, embedded within a larger biomolecular spin network containing several proton spins. We choose the butterfly operator to be a local Pauli-$X$ operator acting on carbon $b$, while the measurement operator acts on carbon $m$,
\begin{align}
    M &= X_{m} = M^{\dagger} \\
    B(t) &= e^{iH_{\mathrm{eff}}t}X_{b} e^{-iH_{\mathrm{eff}}t}
        = B^{\dagger}(t).
\end{align}
The two carbon atoms therefore serve as well-defined input and probe sites within the larger interacting spin network. Although $B(0) = X_b$ is initially localized on carbon $b$, its subsequent evolution is governed by the full effective Hamiltonian, allowing the operator to spread through the intervening proton network and eventually acquire support on carbon $m$.

\subsection{Pipeline for modeling OTOCs in protein systems}\label{sec:pipeline}
\begin{figure*}
    \centering
    \includegraphics[scale=0.5]{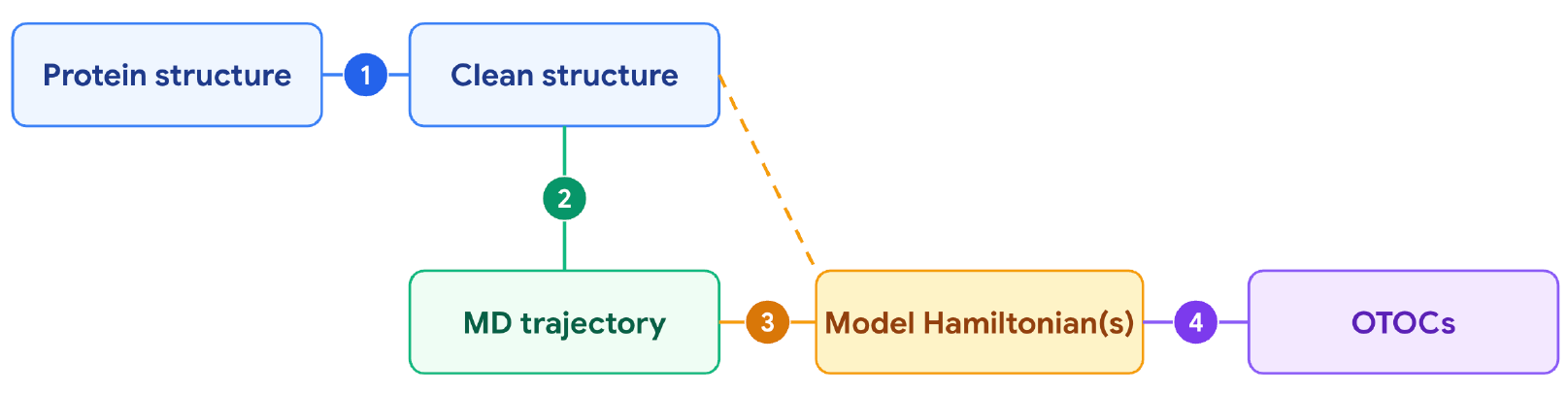}
    \caption{A schematic of the OTOC simulation pipeline used in this work.}
    \label{fig:pipeline}
\end{figure*}

One of the primary goals of this work is to describe a pipeline for simulating OTOCs in protein systems. This is relevant both as a means of understanding the properties of these objects as implemented in real experiments and also as part of the structure fitting process (see Sec.~\ref{sec:model_fitting_structure}). The procedure starts from a model structure. This can be a partial/incomplete model, a lower-resolution structure, or an AlphaFold model as discussed in Sec.~\ref{sec:model_fitting_structure}, and then proceeds in 4 steps:
\begin{enumerate}
    \item Cleanup structure: the initial structure likely requires some cleanup (adding protons, setting charge states, missing or undetermined residues, etc).
    \item Run molecular dynamics: molecular dynamics should be used to partially relax the structure and average over fast degrees of freedom. This step is optional, but necessary for good agreement with experiment.
    \item Generate effective Hamiltonians: for each powder average instance, a subset of important spins are selected to yield a finite spin system and the matrix elements of the effective Hamiltonian are computed.
    \item Simulate the OTOC: the OTOC is simulated for each powder instance and the average is computed by contracting with the quadrature weights (Eq.~\eqref{eqn:pa_quad}).
\end{enumerate}
These steps are shown schematically in Fig.~\ref{fig:pipeline}. Steps 1 and 2 will depend on details of the experimental setup and the protein system of interest. More detail on the systems used in this work and how the structures were preprocessed is given in Sec.~\ref{sec:dataset} and Appendix~\ref{sec:App_systems}. The output from step 1 should be a single structure (pdb or mmcif format). The molecular dynamics simulations are described in Sec.~\ref{sec:md}. The output from step 2 is usually a trajectory file (xtc or dcd) with the topology defined by a pdb file. This trajectory can then be downsampled to a smaller number of frames which can be saved as separate models in pdb or mmcif format. In step 3, these frames are used to generate the spin clusters and model Hamiltonians as described in Secs.~\ref{sec:otoc_hamiltonian} and~\ref{sec:otocs_averaging}.

For step 4, we simulate the OTOC using exact state-vector simulation as implemented in the qsim\cite{qsim2025} extension to Cirq\cite{Cirq_Developers_2026}. The OTOCs studied here can be computed from a more symmetric expression,
\begin{equation}
    \mathrm{OTOC} = \frac{1}{2^{n - 1}}\sum_{\mathbf{q}}
    \matrixel{0, \mathbf{q}}{H_mX_b(t) X_mX_b(t)H_m}{0, \mathbf{q}},
\end{equation}
where we have assumed that the measurement spin is the first qubit, $H_m$ is the Hadamard operator acting on the measurement spin, and $\mathbf{q}$ is a bitstring representing the state of the remaining $n - 1$ qubits. We compute the OTOC by sampling over $\mathbf{q}$, evolving the state with the butterfly,
\begin{equation}
    X_b(t)H_m\ket{0,\mathbf{q}},
\end{equation}
and then taking the expectation value of $X_m$. In practice, we sample from a Haar distribution over the final $n - 1$ qubits. This method is exact up to the sampling error and error from Trotterizing the time evolution.

%%%%%%%%%%%%%%%%%%%%%%%%%%%%%%%%%%%%%%%%%
\subsection{The effective Hamiltonian}\label{sec:otoc_hamiltonian}
%%%%%%%%%%%%%%%%%%%%%%%%%%%%%%%%%%%%%%%%%

For simplicity, we consider a solid-state NMR experiment on a static, non-spinning powdered sample (see Appendix~\ref{app:NMR_primer} for a brief primer on the relevant NMR concepts). The resulting dipolar couplings acquire an orientation dependence set by the internuclear vectors relative to the static external magnetic field.  Radio-frequency (rf) pulse sequences then provide the means to shape the effective Hamiltonian. Through average Hamiltonian theory, a cyclic sequence can suppress, select, and tune the relative strengths of individual interactions. For the simulations presented here, we choose the following effective Hamiltonian, which is realizable through a suitable combination of radio-frequency pulses:
\begin{widetext}
\begin{equation}
    H_{\mathrm{eff}} = 2\pi\eta_{\mathrm{CC}}\tilde{b}_{mb}\left[Y^mY^b - X^mX^b\right] + 2\pi\eta_{\mathrm{HH}}\sum_{i < j}\tilde{b}_{ij}
    \left[Y^iY^j - X^iX^j\right] + 2\pi\eta_{\mathrm{CH}}\sum_{p}\sum_{i}
    \tilde{b}_{pi}Z^iZ^p,
    \label{target}
\end{equation}
\begin{equation}
    \tilde{b}_{X} \equiv b_{X}d^2_{00}(\beta_{X}^{\rm L}) = 
    b_{X}\frac{1}{2}\left(3\cos^2\beta^{\mathrm {L}}_{X} - 1\right) \qquad [X = (m,b), (i,j), (p,i)],
    \label{dipolar_coupling}
\end{equation}
\end{widetext}
where $\beta^{\mathrm{L}}_X$ is the angle between the internuclear spin vector involved and the static magnetic field, $\vec{\mathrm{B}}_0$ - the laboratory frame (L). Here, $(i,j)$ index the spins of the $^1$H bath, and $(m,b)$ index the $^{13}$C measurement and butterfly spins respectively. The corresponding dipolar coupling constants, defined in Hz, are
\begin{align}
    b_{mb} &\equiv -\frac{1}{2\pi}\frac{\mu_0}{4\pi}\frac{\gamma_C^2 \hbar}{r_{mb}^3}, \label{eq:dipolar_coupling_constants1}\\
    b_{ij} &\equiv -\frac{1}{2\pi}\frac{\mu_0}{4\pi}\frac{\gamma_H^2 \hbar}{r_{ij}^3}, \label{eq:dipolar_coupling_constants2}\\
    b_{pi} &\equiv -\frac{1}{2\pi}\frac{\mu_0}{4\pi}\frac{\gamma_C\gamma_H \hbar}{r_{pi}^3}.
    \label{eq:dipolar_coupling_constants3}
\end{align}
The parameters $\eta_{\mathrm{CC}}$, $\eta_{\mathrm{HH}}$, and $\eta_{\mathrm{CH}}$ represent scaling factors arising from Hamiltonian engineering. The bounds on these parameters define the space of effective Hamiltonians accessible through rf sequence design, as detailed in Appendix~\ref{app:Ham_design} and Appendix~\ref{sec:scaling factor bounds}.

Eq.~\eqref{dipolar_coupling} applies to static solid samples. Magic-angle spinning (MAS) is commonly employed in solid-state NMR to remove large chemical shifts and dipolar anisotropy that hinders interpretation of results. Here, the sample is mechanically rotated about an axis oriented at the magic angle $\beta_{\rm ma} = \arccos(1/\sqrt{3})$ to the static magnetic field. Under these conditions, the same operator structure can be recovered using an appropriate recoupling sequence that reintroduces the desired interactions~\cite{doi:https://doi.org/10.1002/9780470034590.emrstm0551}. The exact design principles are detailed in Appendix~\ref{app:NMR_primer}, and the structure of Eq.~\eqref{target} can be reproduced under MAS, with new dipolar coupling constants given by 
\begin{equation}
    \tilde{b}_X = 2b_Xd^2_{m'0}(\beta_{\rm ma})d^2_{m'0}(\beta^{\rm R}_{X})\cos\{m'(\gamma^{\rm R}_{X}-\alpha^0_{\rm RL})\},
    \label{eq:mas_dipolar_couplings}
\end{equation}
where $(\beta_X^{\rm{R}}, \gamma_X^{\rm R})$ are the polar and azimuthal angles of the internuclear spin vector in the rotor frame, $m'$ can be chosen as 1 or 2, and $\alpha^0_{\rm RL}$ describes the starting position of the rotor (see Appendix \ref{app:NMR_primer} for more detail).

The choice of our target Hamiltonian is motivated by two complementary considerations. From the NMR perspective, $H_{\rm eff}$ represents a double-quantum (DQ) Hamiltonian that can be efficiently engineered using radio-frequency pulse sequences, and its sign can be readily inverted by appropriate control of the pulse sequence. From the quantum-information perspective, $H_{\rm eff}$ provides a natural setting to study operator spreading and scrambling, which are central to the dynamics probed by OTOCs. For the simulations presented here, we choose a symmetric Hamiltonian design, motivated by experimentally established pulse-sequence protocols~\cite{zhang2025quantum}
\begin{equation}
    \eta_{\mathrm{CC}} = \eta_{\mathrm{HH}} = 0.15, \qquad \eta_{\mathrm{CH}} = 0.3.
    \label{optimal sfs}
\end{equation}

However, as discussed in Appendix~\ref{app:general_eff_DQHam}, other choices within the accessible space can lead to faster OTOC dynamics, with potential advantages in the presence of decoherence.

%%%%%%%%%%%%%%%%%%%%%%%%%%%%%%%%%%%%%%%%%
\subsection{Averaging over molecular motion and orientation}\label{sec:otocs_averaging}
%%%%%%%%%%%%%%%%%%%%%%%%%%%%%%%%%%%%%%%%%
In real experiments, typically two statistical averages have to be considered. The first arises from molecular motion. Even in a solid, molecules undergo residual motion arising from processes such as vibrations, local rotations, and conformational fluctuations. These motions span a broad range of timescales that can be classified relative to the timescale of the spin dynamics. When a molecular motion occurs on a timescale much shorter than that of the spin dynamics, the spin system effectively experiences an average over the corresponding molecular configurations. We can therefore define an effective spin interaction by averaging over the fast molecular degrees of freedom
\begin{equation}
    \tilde{b}_X(\hat{\mathbf{n}}) \equiv \langle \tilde{b}_X(\beta_X)\rangle,
\end{equation}
where $\hat{\mathbf{n}}$ denotes the direction of the molecular frame axis relative to the laboratory frame, i.e., the orientation of the static magnetic field, $\vec{\mathrm B}_0$. A simple estimate can be obtained by evaluating the coupling for a single static molecular structure
\begin{equation}
   \langle \tilde{b}_X(\beta_X)\rangle \approx \tilde{b}_X(\beta_X^{\mathrm{static}}),
\end{equation}
which neglects the detailed effects of molecular motion. Here, we instead account for these effects by sampling molecular dynamics over an early-time trajectory and averaging the coupling over the sampled configurations,
\begin{equation}
   \langle \tilde{b}_X(\beta_X)\rangle \approx \langle \tilde{b}_X(\beta_X)\rangle_{\mathrm{MD}},
\end{equation}
details of the MD simulations are given in Sec.~\ref{sec:md}.

Second, the sample may contain molecules with different overall orientations relative to the laboratory frame (see Appendix~\ref{app:NMR_primer}). This is particularly relevant for powdered samples, where the molecules have no preferred orientation. The observed OTOC signal is therefore an ensemble average over the OTOCs associated with the different molecular orientations present in the sample. Since the overall molecular orientation is typically static on the timescale of the experiment, this produces a distribution of $\beta_X$ values across the ensemble. For the Hamiltonian considered here, the orientational average reduces to an integral over the two-dimensional space of orientations of the molecular axis relative to the laboratory frame:
\begin{equation}
    \langle\mathrm{OTOC}\rangle_{\mathrm{PA}} = \frac{1}{4\pi}\int_{S^2} d\hat{\mathbf{n}} \mathrm{OTOC}(\hat{\mathbf{n}}).
\end{equation}
In practice, we compute this integral by quadrature, discretizing over a Lebedev grid~\cite{LEBEDEV197610},
\begin{equation}\label{eqn:pa_quad}
    \langle\mathrm{OTOC}\rangle_{\mathrm{PA}} \approx \sum_{i}w_i \mathrm{OTOC}(\hat{\mathbf{n}}_i).
\end{equation}
Both the couplings and the Lebedev grid have inversion symmetry which means that using such a grid allows us to evaluate the OTOC at half as many points (e.g. the 26-point quadrature rule can be evaluated with only 13 points).

Under magic-angle spinning, the molecular orientation must instead be specified relative to the rotor frame, which is subsequently transformed into the laboratory frame. The rotation of the rotor breaks the axial symmetry about the laboratory $z$ axis exploited in the static case, so the powder average generally retains a dependence on two orientational angles ($\beta_{X}$, $\gamma_{X}$).

%%%%%%%%%%%%%%%%%%%%%%%%%%%%%%%%%%%%%%%%%
\subsection{Spin network truncation}\label{sec:spin_network_truncation}
%%%%%%%%%%%%%%%%%%%%%%%%%%%%%%%%%%%%%%%%%

OTOCs, being a short-time experiment, are only sensitive to spins around the measurement operator $m$, and the butterfly operator $b$.
This follows from the concept of an ``operator front''~\cite{vonkeyserlingkOperatorHydrodynamicsOTOCs2018a,nahumOperatorSpreadingRandom2018a}: a local spin polarization $M$ spreads out in time to those operators nearby, and after some time $t$ will only have non-negligible support on operators within some finite volume centered on its initial position $\mathbf{r}_m$.
We call the edge of this volume the operator front.
The presence of any spin outside this front is invisible to any measurements on the operator $M(t)$.
This implies we can remove these spins from a simulation without affecting the final result; we call this process truncation.
In the absence of error, an OTOC allows even greater truncation: we can remove a spin unless it lies in the intersection between the front of the measurement operator $M$ and the front of the butterfly operator $B$.
This allows us to reduce a spin network of thousands or more spins to one of tens or hundreds, a significant saving.

For large protein or solid-state systems, truncating the spin network to include only spins relevant to the OTOC is essential to make the problem tractable on quantum hardware.
(This was not an issue for Ref.~\cite{zhang2025quantum} as the systems contained only $9$ and $15$ spins; all were found to be relevant to the OTOC.)
However, determining whether a given spin is relevant to the OTOC is challenging, due to both the cost of estimating the error from removing a spin, and the number of possible combinations of spins to keep.
There are many potential heuristics one can consider to make the calculation tractable; in this work we propose a two-step solution.
First we calculate a semi-classical approximation $r(i)$ to the ``relevance'' of spin $i$ to the OTOC (ideally related to the error incurred in removing $i$). Due to the dependence of the couplings on molecular orientation, this relevance ranking must be done separately for each orientation on the powder average.
Then, we add spins one at a time to $H_{\mathrm{eff}}$, choosing for each orientation the most relevant remaining spin, so that each orientation contains the same number of spins. The powder averaged OTOC is re-simulated after each addition, until we see convergence. 
We take convergence to mean the OTOC remaining within some error bound $\epsilon$ as further spins are added.
Optimizing this heuristic or finding more optimal choices are clear tasks for future work.

The performance of our two-step heuristic depends heavily on the ranking of the spins; we propose one such ranking here.
Motivated by the above observation that information has to pass from $m$ to $b$ through a spin $i$ for $i$ to be relevant, we take a semi-classical approximation to the time-of-flight of $m\rightarrow i\rightarrow b$ that targets being simple to calculate.
We define the matrix $A_{\mathrm{eff}}$ as the element-wise inverse of the dipolar coefficient matrix that gives $H_{\mathrm{eff}}$:
\begin{align}
    [A_{\mathrm{eff}}]_{p,q} &= |2\pi\eta_{\mathrm{CC}}\tilde{b}_{pq}|^{-1} \\
    [A_{\mathrm{eff}}]_{p,i} &= |2\pi\eta_{\mathrm{CH}}\tilde{b}_{pi}|^{-1} \\
    [A_{\mathrm{eff}}]_{i,j} &=|2\pi\eta_{\mathrm{HH}}\tilde{b}_{ij}|^{-1}. 
\end{align}
Elements in $A_{\mathrm{eff}}$ are proportional to the time it would take for an operator to spread from site $i$ to site $j$ if all other couplings were turned off.
$A_{\mathrm{eff}}$ is then the adjacency matrix of a weighted graph; we can calculate the length $l(i,p)$ of the shortest path on this graph from all spins $i$ to $p=m$ and $p=b$ using Dijkstra's algorithm.
The length $l(i,m) + l(i,b)$ is a poor man's classical approximation for the time required for an operator to spread from $m\rightarrow i\rightarrow b$ (up to a multiplicative factor).
As paths are more relevant the shorter this time is, we estimate the relevance as
\begin{equation}
    r(i) = \big(l(i,m) + l(i,b)\big)^{-1}.
\end{equation}
This yields a metric that is a balance between being accurate at ranking spins, and efficient to classically compute.

\section{Protein Structure Dataset}\label{sec:dataset}
\begin{figure*}
    \centering
    \includegraphics[scale=0.5]{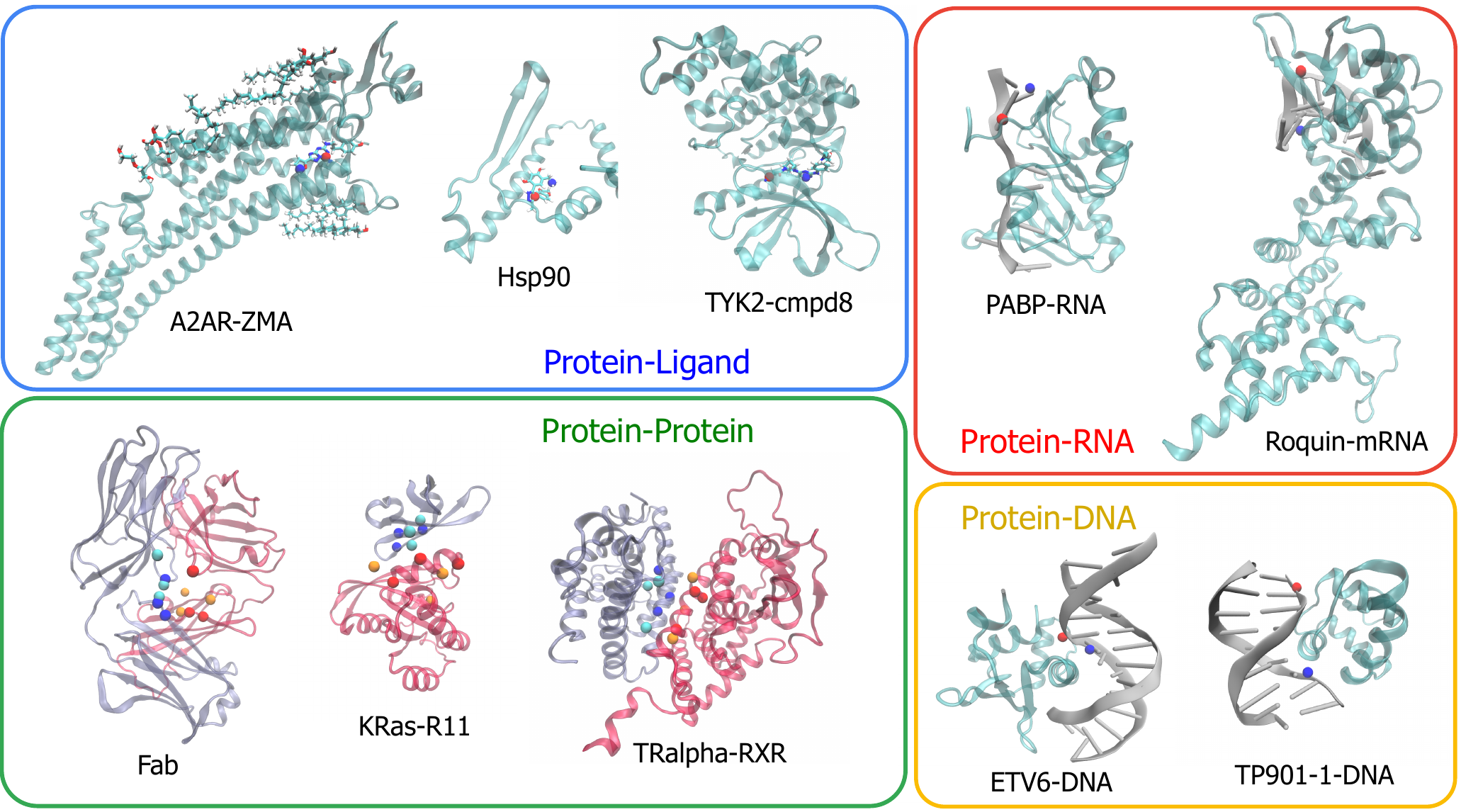}
    \caption{Ribbon structures of all 10 protein systems with the locations of important carbons tagged. For the protein-ligand systems, blue (red) spheres indicate the central (distal) ligand carbons. For the protein-protein systems the two proteins are distinguished by red and blue ribbons and dark blue (cyan) or red (orange) spheres are used to show the alpha (side-chain) carbons. For the protein-DNA/RNA systems blue (red) spheres show the locations of the nucleobase (sugar) carbons.}
    \label{fig:proteins}
\end{figure*}

Our dataset is constructed of examples meant to be representative of the following three common problems in protein structure determination:
\begin{enumerate}
    \item Protein-ligand binding
    \item Protein-protein interactions and quaternary structure
    \item Structure of DNA/RNA bound to proteins
\end{enumerate}
In Appendix~\ref{sec:App_systems} we describe each protein system, provide the reference to the original structure, and describe how the structure was processed. In Fig.~\ref{fig:proteins}, we show the structures with the locations of some of the relevant carbons tagged.

\subsection{Selection of carbon-carbon pairs}
For each system we selected a set of 30 - 40 carbon-carbon pairs with a C-C distance between 5 - 17~\AA. Furthermore, we wanted pairs of atoms where the OTOC is likely to be sensitive to the important structural information, but with different specific sites (for example backbone, or side-chain) represented. We do not consider how easy it is to label particular sites, so it may be that certain pairs of carbon-13 atoms may not be realizable in practice.

\subsubsection{Protein-ligand systems}
For the protein-ligand systems we choose one ``central'' carbon atom near the geometric middle of the ligand and one ``distal'' carbon atom close to the edge of the ligand. For each ligand carbon we then construct two series of carbon-carbon pairs: one using a series of protein alpha carbons and one using a series of protein side-chain carbons. In total we have 4 series or approximately 40 C-C pairs for each system. We label each series as
\begin{itemize}
    \item ``central:alpha'' (CA): Central ligand carbon with alpha carbons of the protein
    \item ``central:side-chain'' (CS): Central ligand carbon with side-chain carbons of the protein
    \item ``distal:alpha'' (DA): Distal ligand carbon with alpha carbons of the protein
    \item ``distal:side-chain'' (DS): Distal ligand carbon with side-chain carbons of the protein
\end{itemize}
Each series is chosen so that the C-C distances are between 5 - 15~\AA{} in the initial PDB structure. Each ligand carbon has its own series of alpha or side-chain carbons and these atoms are chosen randomly such that the distances span the target distance uniformly.

\subsubsection{Protein-DNA/RNA systems}
For the protein-DNA/RNA systems we choose one carbon on the nucleobase and one carbon atom on the sugar backbone. For each DNA/RNA carbon atom we then construct two series of carbon-carbon pairs: one using a series of protein alpha carbons and one using a series of protein side-chain carbons. In total we have 4 series or approximately 40 C-C pairs for each system. We label each series as
\begin{itemize}
    \item ``nucleobase:alpha'' (NA): Nucleobase carbon with alpha carbons of the protein
    \item ``nucleobase:side-chain'' (NS): Nucleobase carbon with side-chain carbons of the protein
    \item ``sugar:alpha'' (SA): Sugar carbon with alpha carbons of the protein
    \item ``sugar:side-chain'' (SS): Sugar carbon with side-chain carbons of the protein
\end{itemize}
Each series is chosen so that the C-C distances are between 5 - 15~\AA{} in the initial PDB structure. Each nucleobase/sugar carbon has its own series of alpha or side-chain carbons and these atoms are chosen randomly such that the distances span the target distance range uniformly. In practice there are somewhat fewer than 40 pairs for each system since some distances may not exist in a given structure.

\subsubsection{Protein-protein systems}
For the protein-protein systems we assume that the interesting structural information involves the relative orientation of the two proteins. We therefore focus on the protein-protein interface where we choose 3 alpha carbon atoms on each protein and 3 side-chain carbon atoms on each protein. Restricting to pairs where one carbon is on each protein, this gives us 4 series of C-C pairs which we label as
\begin{itemize}
    \item ``alpha:alpha'' (AA): An alpha carbon on the first protein and an alpha carbon on the second
    \item ``alpha:side-chain'' (AS): An alpha carbon on the first protein and a side-chain carbon on the second
    \item ``side-chain:alpha'' (SA): A side-chain carbon on the first protein and an alpha carbon on the second
    \item ``side-chain:side-chain'' (SS): An side-chain carbon on the first protein and a side-chain carbon on the second
\end{itemize}
This would provide a total of 36 C-C pairs for each system except that we exclude pairs with a distance greater than 17~\AA, so the actual number is lower in practice.

\subsection{Molecular dynamics simulations}\label{sec:md}
Molecular dynamics (MD) simulations were used to account for averaging over the fast molecular degrees of freedom. The precise form of this average will depend on the experimental details including the type of solid, the temperature, and if the solvent is deuterated. For the purpose of this study, we are not targeting any specific experimental setup, but we want to include the qualitative effect of the fast degrees of freedom. For this reason, we chose to use solution-phase molecular dynamics with backbone restraints so that the molecular motion is more constrained as it would be in the solid state. The choice of water as solvent is somewhat arbitrary, but, since the force fields used here are primarily designed to be used with a specific water model, this is the simplest choice. We do not expect the choice of solvent to qualitatively change the fast degrees of freedom that we aim to sample with MD.

All initial PDB files were obtained from the Protein Data Bank (PDB) entry for each system~\cite{proteindatabank}. These PDB files were then cleaned with QSimulate's ProtClean program~\cite{protclean} which standardizes PDB files by adding missing hydrogens, loops, and sidechains, determining protonation states of residues, and standardizing residue IDs and atom numbering.

For molecular dynamics simulations, OpenMM 8.5~\cite{eastman2023openmm} was used to solvate each system with water using a 10~\AA{} buffer region. The Amber ff14SB force field~\cite{maier2015ff14sb} included with OpenMM was used for proteins and the Amber OL15~\cite{zgarbova2011ol15} and OL3~\cite{zgarbova2011ol3} force fields were used for DNA and RNA strands, respectively. Small molecules, ligands, and noncanonical residues were parameterized with GAFF 2.2~\cite{wang2004gaff, wang2006automatic} using the OpenMMForceFields package~\cite{chodera2026openmmforcefields}. TIP3P~\cite{jorgensen1983tip3p} was used for the water model.
To constrain the protein structures, a periodic torsion force of 100 kJ/mol was added with a periodicity of 1 to the initial angles $\phi$ and $\psi$ of each residue. No constraints were added to the DNA/RNA sugar-phosphate backbones or the ligands. Each system was then minimized and equilibrated for 20 ps at 300 K with a timestep of 2 fs in the NVT ensemble at the initial box dimensions. NPT relaxation was not performed but production run densities were checked for the absence of any negative pressure effects. Following equilibration, a production run of 2 ns at 300 K in the NVT ensemble was performed with frames saved every 5000 steps. All MD simulations were performed using nonbonded contributions calculated with PME and a cutoff of 10~\AA{} and constrained hydrogen atoms. The saved MD trajectory was aligned to the first frame to remove any residual translation/rotation, downsampled to 100 frames, and saved in .pdb format. These 100 frames were used to approximate the Hamiltonian averaged over the fast degrees of freedom.

\begin{figure*}
    \centering
    \includegraphics[scale=0.7]{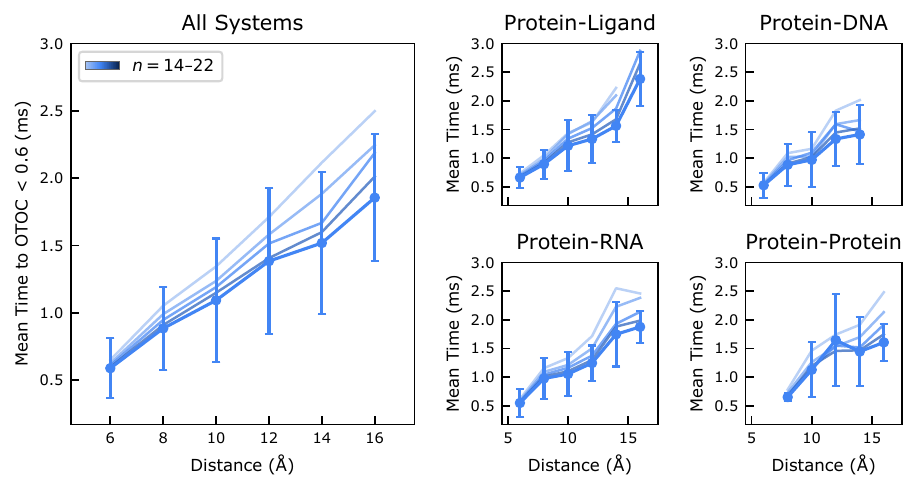}
    \caption{The mean distance dependence of the operator front for all systems and broken down by system type. The results are binned by distance and shown for smaller spin network sizes (lighter lines). Any example where the OTOC does not fall below 0.5 over the computed timescale is excluded. The systems are binned by mean distance and any bin with fewer than 5 examples is excluded.}
    \label{fig:mean_operator_front}
\end{figure*}

\subsection{Dataset size and composition}
The full dataset consists of the 10 systems described in Appendix~\ref{sec:App_systems}, and in total it includes 367 carbon-carbon pairs with average distances from molecular dynamics between 4.67~\AA{} and 17.92~\AA{}. While the atom pairs were chosen from the static structures to span a distance of 5 - 17~\AA{}, this distance can fluctuate during the MD simulation, and the average can be different from the distance in the static structure. For these systems, the OTOC was computed for $t = 0 - 3.7$ ms for $n = 14 -22$ spins. For 13 of 367 pairs, the OTOC did not decay below 0.5 by the end of the simulated time window. These instances are omitted from the subsequent analysis, leaving 354 OTOCs.

A subset of 4 systems (ETV6-DNA, Hsp90, Fab, Roquin-mRNA) were used to evaluate the effects of the different types of averaging. We will refer to the OTOCs for these systems as dataset-A which includes 150 C-C pairs in total. For these systems we additionally computed the OTOCs from the static crystal structure for $n = 14 - 22$ spins. For the static systems the OTOC was computed for $t = 0 - 1.9$ ms. For 12 of 150 pairs, the OTOC did not decay below 0.5 by the end of the simulated time window. These instances are omitted from the subsequent analysis, leaving 138 OTOCs in dataset-A. 

A small subset of 8 specific C-C pairs was used to evaluate approximate classical simulation methods. For this dataset-C we use the following C-C pairs with the given average C-C distances:
\begin{itemize}
    \item ETV6-DNA (NS-04): 9.37~\AA{}
    \item ETV6-DNA (SA-07): 12.51~\AA{}
    \item Hsp90 (CS-00): 5.41~\AA{}
    \item Hsp90 (DA-08): 15.41~\AA{}
    \item Fab (AA-06): 9.58~\AA{}
    \item Fab (SS-07):  14.52~\AA{}
    \item Roquin-mRNA (NA-00): 6.05~\AA{}
    \item Roquin-mRNA (SS-08): 12.85~\AA{}
\end{itemize}
These systems were chosen randomly to span the relevant distances while requiring (1) exactly 2 pairs chosen from each system in dataset-A, (2) all pairs coming from distinct series, and (3) all OTOCs decaying to below 0.5 by the final time for the $n = 22$ calculation.

\begin{figure*}
    \centering
    \includegraphics[scale=0.7]{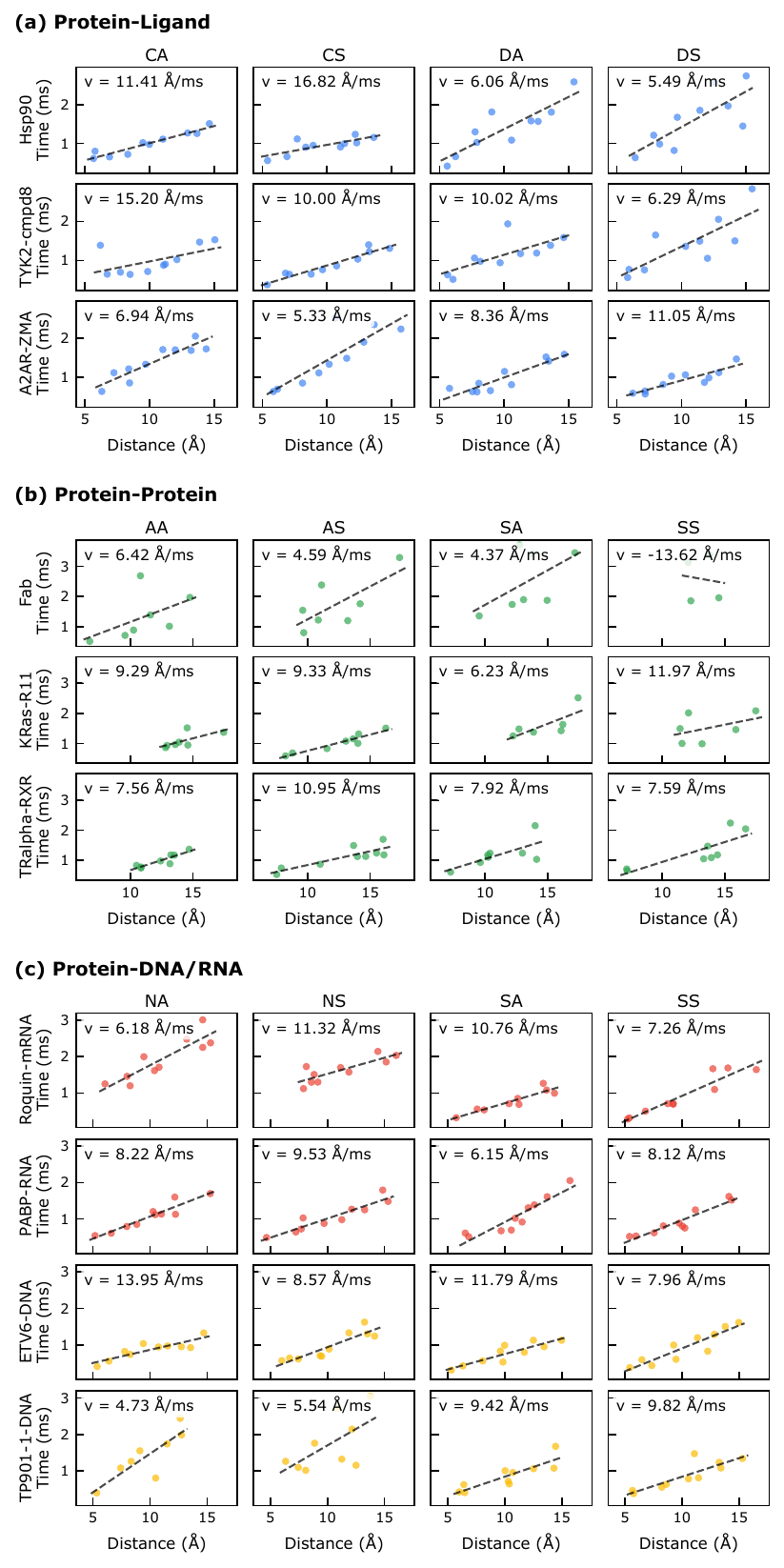}
    \caption{The behavior of the operator front broken down by pair type for all 10 systems. With the possible exception of the protein-protein systems, the behavior is approximately linear so we can extract an operator velocity (Eq.~\eqref{eqn:operator_velocity}). There is one negative velocity for a series (Fab: SS) where there are not enough points to get a reliable fit. Note that the fits also yield an intercept which ranges from about $-0.93$ to $0.64$ ms (discounting the Fab: SS with negative velocity) and is usually close to zero ($0 \pm 0.5$ ms).}
    \label{fig:operator_front_all}
\end{figure*}

\section{Results: OTOCs in protein systems}\label{sec:results_otocs}
In Sec.~\ref{sec:results_information} we analyze the information propagation and the sensitivity of the OTOC to structure over the entire dataset. In Secs.~\ref{sec:results_spin_network} and~\ref{sec:results_averaging} we discuss two important practical details: truncation of the proton spin network and simulation of averaging introduced in Sec.~\ref{sec:otocs_averaging}.

\subsection{Information propagation and sensitivity to structure}\label{sec:results_information}
In Fig.~\ref{fig:mean_operator_front} we bin all systems by C-C distance and show how the time for the OTOC to fall to 0.6 varies as a function of distance. This ``operator front'' propagates outward with time, but it is difficult to say anything more quantitative about the behavior on average, even with the results split by system type. In the protein-protein systems the dependence on distance is less pronounced than in the other systems. This is to be expected because with the measurement and butterfly spins on separate proteins, information propagating through the proton spin network has to flow through relatively few close contacts. So for these systems, any reasonable notion of distance in the proton network may not correlate with distance in space. This idea is echoed in Fig.~\ref{fig:operator_front_all} where we show results broken down by specific systems and pair types. When broken down in this way, an empirically linear relationship between distance and time is observed for all systems with the exception of some of the protein-protein examples. From a linear fit we can extract an operator velocity,
\begin{equation}\label{eqn:operator_velocity}
    v = \frac{\text{distance}}{\text{arrival time}},
\end{equation}
which ranges from 4.37 - 16.82~\AA/ms for these systems. Here we define the ``arrival time'' to be the time where the OTOC first falls below 0.6. There is no reason to expect a linear relationship in general, and previous work suggests that, for 3-dimensional systems with interactions that decay with $r^{-3}$, the operator growth should be superpolynomial\cite{chen2019quantum,zhou2020operator,zhou2023operator}. This implies that we should expect a sublinear ``arrival time'' when we plot time vs distance. The fact that we observe a linear relationship could be attributed in part to the results not being fully converged with respect to the size of the spin network (see Sec.~\ref{sec:results_spin_network}). Given that the unconverged spin truncation may artificially linearize the operator velocity, the extracted velocities could be interpreted as lower bounds to the true value. Also, the superpolynomial operator growth may not be apparent over small distances. This is consistent with previous work\cite{zhou2023operator}, and, for these protein systems in particular, can be rationalized by noting that the spin network may have an effective spatial dimensionality of less than three over short distances. Nevertheless, these velocities provide a rough estimate of the timescale for information propagation in these systems over the distances studied here. Furthermore, we can use the velocity range to estimate the time needed to perform experiments. For a distance of 15~\AA, the arrival time is between 0.89 and 3.43 ms. We can multiply this range by a factor of 1.5 to account for running the experiment past the arrival time and by an additional factor of 2 to account for forward and reverse propagation. This yields an estimate of between 2.67 and 10.29  ms in total.

\begin{figure*}
    \centering
    \includegraphics[scale=0.7]{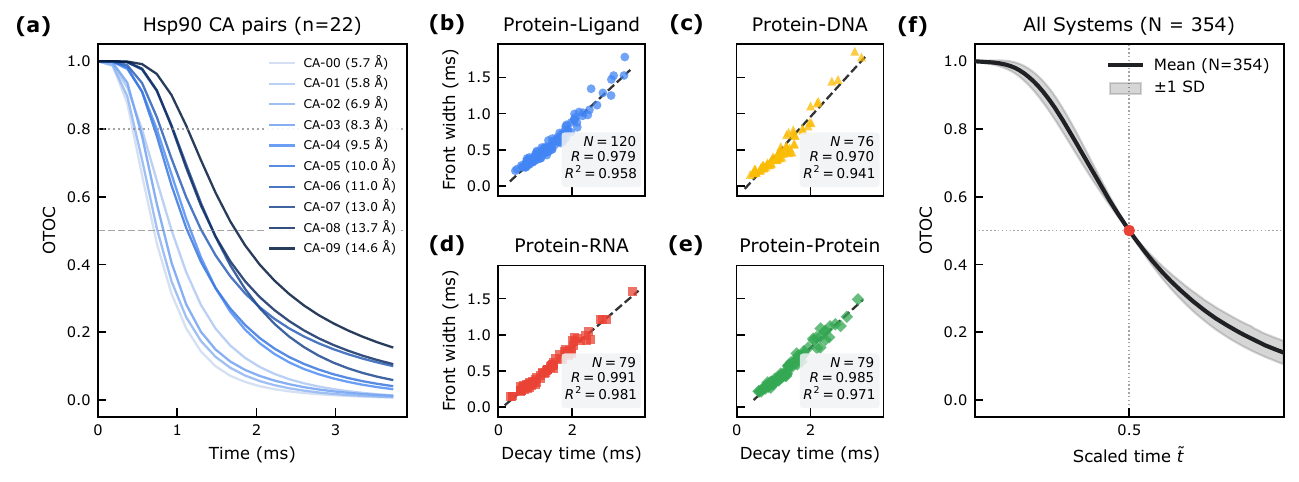}
    \caption{(a) An example of how the OTOCs depend on distance for a particular system. (b) - (e) Correlation between the operator front width, defined as the time to decay from 0.8 to 0.5, and the total decay time. (f) The mean and standard deviation for OTOCs of all systems scaled by their decay time.}
    \label{fig:operator_front_width}
\end{figure*}
In Fig.~\ref{fig:operator_front_width}(a) we show the OTOC curves for a series of C-C pairs in one particular system, and in Fig.~\ref{fig:operator_front_width}(b) - (e) we show that the width of the OTOC decay is linearly correlated with total decay time. This suggests that the shape of the OTOC is relatively constant among different systems, a point which is emphasized in Fig~\ref{fig:operator_front_width}(f). This already suggests that only a single structural parameter can be extracted from the OTOC for a given pair of atoms because the OTOC curve is almost entirely described by a single parameter: the decay time.

\begin{figure}
    \centering
    \includegraphics[scale=0.7]{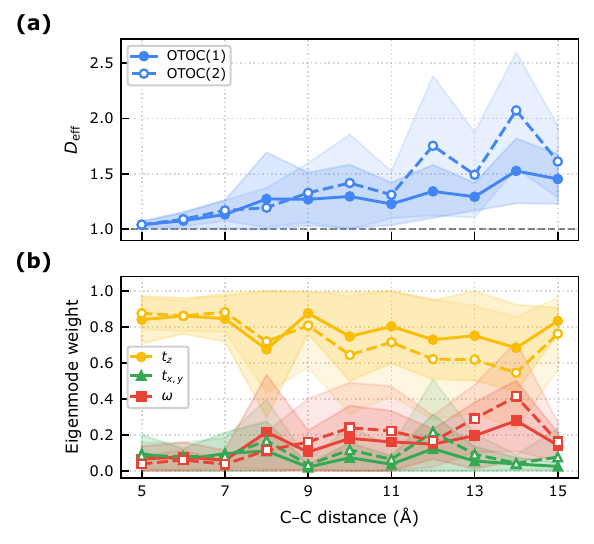}
    \caption{(a) Effective dimension $D_{\rm eff}$ as a function of the distance between the measurement and butterfly $^{13}$C spins for different correlators on the 12-spin proxy system. The solid/dashed (OTOC(1)/OTOC(2)) lines and shaded regions represent the mean and standard deviation over 10 random proton-bath geometries, respectively. (b) Weights of the translational (split into $|v_{t_z}|^2$ and $|v_{t_x}|^2+|v_{t_y}|^2$) and rotational ($|v_{w_x}|^2+ |v_{w_y}|^2+ |v_{w_z}|^2$) components of the dominant eigenvector ($v$), characterizing the structural degrees of freedom to which OTOC(1) and OTOC(2) are most sensitive. The dominance of translation along the internuclear vector ($t_z$) for both correlators suggests that the interspin distance is the primary structural parameter accessible from a single OTOC trajectory in this proxy model.}
    \label{fig:FIM}
\end{figure}

To investigate the nature of the structural information encoded by a single OTOC, we perform an additional analysis using a simplified 12-spin proxy for the local protein–ligand environment, consisting of two methyl groups separated by a bath of protons (Appendix~\ref{sec:toy_model}). Specifically, we examine the sensitivity of the OTOC to rigid translations and rotations of the ligand relative to its surrounding proton environment, 
Specifically, we examine the sensitivity of the OTOC to rigid translations and rotations of the ligand relative to its surrounding proton environment through the sensitivity metric $J^\mathsf{T}J$, where $J$ is the Jacobian of the OTOC with respect to geometric perturbations (see Appendix~\ref{sec:toy_model} for details). The eigenspectrum characterizes the relative sensitivity of the OTOC to different combinations of structural parameters. To quantify how many of these combinations contribute appreciably to the OTOC dynamics, we construct the eigenmode participation ratio from the eigenvalues, $\lambda_k$,
\begin{align}
\mathrm{D_{\rm eff}} = \frac{(\sum_k \lambda_k)^2}{\sum_k \lambda_k^2}.
\label{eq:effective_dimension}
\end{align}
This quantifies how many independent structural parameter combinations contribute to changes in the OTOC trajectory. Fig.~\ref{fig:FIM} tracks the mean and standard deviation of this effective dimension over typical  $^{13}$C-$^{13}$C distances, for 10 random proton bath geometries. The consistently low effective dimensionality, arising from a single dominant eigenvalue, indicates that the OTOC is primarily sensitive to one dominant combination of structural parameters, even when the underlying structural model contains multiple degrees of freedom. Examining the corresponding eigenmode reveals that its largest contribution comes from translations along the internuclear vector connecting the measurement and butterfly carbons ($t_z$), rather than transverse translations ($t_x, t_y$) or rotations. Together, these results suggest that, for the systems considered here, the interspin distance is the dominant structural parameter that can be reliably extracted from a single OTOC trajectory. It is possible that for larger polyatomic ligand systems, with denser proton baths, OTOCs may display a more multi-modal sensitivity.

Alternatively, recent work~\cite{google25constructive} has suggested that higher-order correlators of the form
\begin{equation}
\mathrm{OTOC}(k) = \frac{1}{2^n}\mathrm{Tr}\left[(B(t)M)^{2k}\right],
\label{eq:otoc(k)}
\end{equation}
may exhibit increased sensitivity to perturbations with increasing $k$, where $\mathrm{OTOC}(1)$ is the correlator primarily investigated in this work. Motivated by these possible advantages, we repeat the above analysis for the second order correlator OTOC(2), requiring two forward-backward time evolutions on the state vector. As demonstrated in Fig.~\ref{fig:FIM}, the qualitative conclusions remain the same - with the OTOC(2) similarly dominated by the $t_z$ mode, with only a modestly larger $D_{\rm eff}$ than OTOC(1) for $r_{\rm CC} \geq 10$~\AA. A more systematic investigation of higher-order correlators, however, is beyond the scope of the present work.

\subsection{Spin network size}\label{sec:results_spin_network}
\begin{figure}
    \centering
    \includegraphics[scale=0.7]{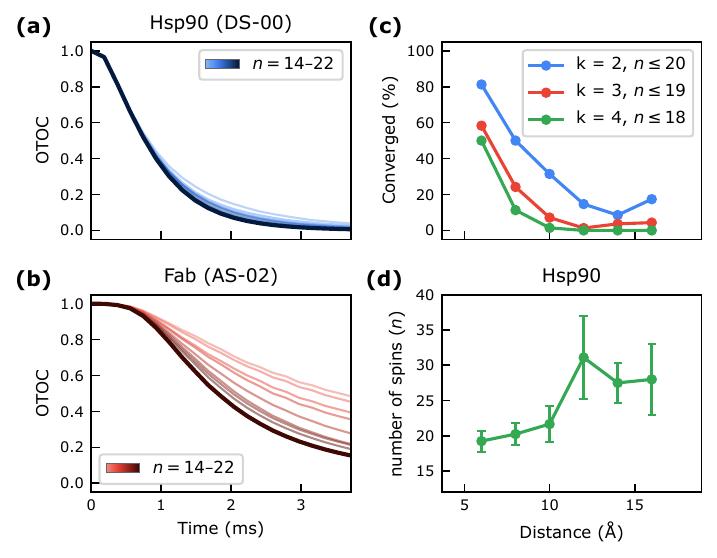}
    \caption{(a) - (b) Two examples of OTOC convergence as a function of spin network size. (a) An example OTOC that is likely converged with respect to spin network size (C-C distance: $6.6$~\AA). (b) An example OTOC that is not converged with respect to spin network size (carbon-carbon distance: $9.0$~\AA). (c) Percentage of OTOCs converged as a function of distance. All carbon-carbon distances are binned (bin width of 2~\AA). $k = 2$ means that $n = 20, 21$ are within $\epsilon = 0.02$ of the reference $n = 22$ curve, $k = 3$ means that $n = 19, 20, 21$ are within the $\epsilon$ of the reference curve and $k = 4$ means that $n = 18, \ldots, 21$ are within $\epsilon$ of the reference curve. (d) Estimated number of spins needed for convergence based on adding each of the 50 - 18 = 32 spins from the $n = 50$ cluster to the core $n = 18$ spin system for all Hsp90 systems in a one-by-one fashion.}
    \label{fig:spin_convergence}
\end{figure}

We described in Sec.~\ref{sec:spin_network_truncation} that OTOCs are insensitive to spins that lie far from the path between the butterfly and measurement spin.
However, it is not at all clear how many spins typically remain in the OTOC as a function of the distance between these spins.
This is a relevant question for the practicality of the OTOC as an analytical tool: systems of up to $30-40$ spins can be simulated immediately with relative ease on classical supercomputers, systems of a few hundred or thousand spins may be simulated in the future on a quantum computer, but yet-larger systems may not be feasible to simulate even on future quantum hardware.
Certainty of convergence is also important to guarantee the other results in this work.
We thus now study the convergence of the OTOC as a function of the number of spins simulated, for different systems and different distances between measurement-butterfly pairs.

For all computed OTOCs, we use the heuristic scheme described in Sec.~\ref{sec:spin_network_truncation} to truncate the proton spin network. 
In Fig.~\ref{fig:spin_convergence}(a) and (b) we show two examples of OTOCs for different spin network sizes: one example where we are confident that the OTOC has converged and one example where the OTOC is not converged. In general, we can never be completely confident of convergence with respect to the size of the truncated spin network, but for a given error tolerance we are increasingly confident of convergence once we see $k$ successive curves within this tolerance ($\epsilon = 0.02$). For example, if we see that all OTOC curves for $n = 18, 19, 20, 21$ are within $\epsilon$ of the reference, $n = 22$, then these $k = 4$ curves provide relatively high confidence that the OTOC is converged at $n = 18$.

In Fig.~\ref{fig:spin_convergence}(c) we show the percentage of carbon-carbon pairs where the OTOC appears to be converged as a function of distance. At short distances ($5 - 7$~\AA) a majority of the $n = 18$ OTOCs are converged with high confidence and it is likely that the vast majority are converged by $n = 22$. However, this number falls rapidly with increasing distance. For carbon-carbon pairs separated by more than 10~\AA, few if any are converged at $n = 18$ and only a small percentage of the OTOCs may be converged at $n = 22$.

To further investigate the necessary spin network size, we computed the matrix elements for large $n = 50$ spin systems for the Hsp90 series of systems. Starting from the $n = 18$ core of each system, we added each of the 32 the remaining spins individually to form a total of 32 distinct $n = 19$ systems. We computed the OTOC for each one so that the number of these systems where adding an extra spin changed the OTOC by more than 0.02 at any time provides a rough estimate of how many spins would be needed for convergence. In Fig.~\ref{fig:spin_convergence}(d) we show this estimate as a function of distance. Estimating the necessary size of the spin network in this way probably yields an underestimate because multiple spins that individually change the OTOC by $< 0.02$ may change the OTOC by more when all included. This suggests that the empirical convergence at $k = 2$ is sometimes unreliable, especially at larger distances. This analysis likely hits the limit of the initial $n = 50$ spin systems around $12 - 17$~\AA{}, but we may nevertheless make a rough guess that a spin network including $n \approx 50 - 100$ spins is necessary to get converged results around distances of $15$~\AA.

\subsection{Averaging over molecular motion and orientation}\label{sec:results_averaging}
\begin{figure}
    \centering
    \includegraphics[scale=0.7]{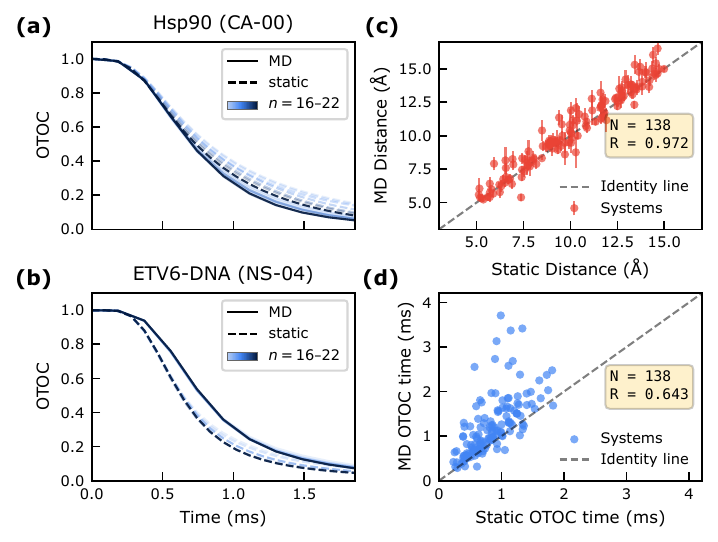}
    \caption{A comparison of the OTOCs from static crystal structures and those from MD averaged structures. (a) An example where the MD averaged OTOC decays faster, and (b) an example where the MD-averaged OTOC decays slower. Convergence with spin network size, $n$, is shown with light colored lines. (c) The correlation between the static and MD C-C distances and (d) the correlation between OTOC decay times. The OTOC decay time is computed as the time it takes the OTOC to decay to 0.6. All results are computed with a 26-point powder average.}
    \label{fig:static_vs_md}
\end{figure}
In Fig.~\ref{fig:static_vs_md} we compare the OTOCs from a Hamiltonian corresponding to the static structure with those from a Hamiltonian corresponding to a short molecular dynamics simulation. Typically, the OTOC with the averaged couplings decays slower indicating that the averaged couplings tend to be weaker. However, this is not always the case and Fig.~\ref{fig:static_vs_md}(a) and (b) show examples of both scenarios. Fig.~\ref{fig:static_vs_md}(c) shows the correlation between the static and MD distances and Fig.~\ref{fig:static_vs_md}(d) shows the correlation between the static and MD OTOC decay times. While the distances are highly correlated between static and MD structures, the timescale for decay in the resulting OTOCs is much more varied. This suggests that the variation in the OTOCs, as shown in Fig.~\ref{fig:static_vs_md}(d), is due to differences in the effective Hamiltonian matrix elements beyond what could be attributed to just slightly shorter or longer distances. This is expected for two reasons. First, the static structure is likely to contain proton-proton distances that are small compared to the average distance in the MD simulation. This is particularly true for static structures derived from crystal structures where the protons are added computationally and the resulting structure is not relaxed. Even though the added protons are added by an algorithm that avoids explicit clashes (unphysically short interatomic distances), the resulting positions may contain inter-proton distances that are shorter than the average. Second, averaging over the fast molecular motion is likely to wash out the most extreme features in the coupling matrix. This is the point of using an MD-averaged effective Hamiltonian, and this analysis suggests that this averaging is necessary. However, the MD simulations used here may include more solvent diffusion and side-chain motion than is typical for solid-state systems. Depending on the details of the experimental setup, these MD simulations may slightly overestimate the effect of averaging over fast motion. We also note that in many cases the OTOCs are not converged with respect to the spin network size at $n = 22$ spins. This could further lead us to overestimate the effect of MD averaging.

\begin{figure}
    \centering
    \includegraphics[scale=0.7]{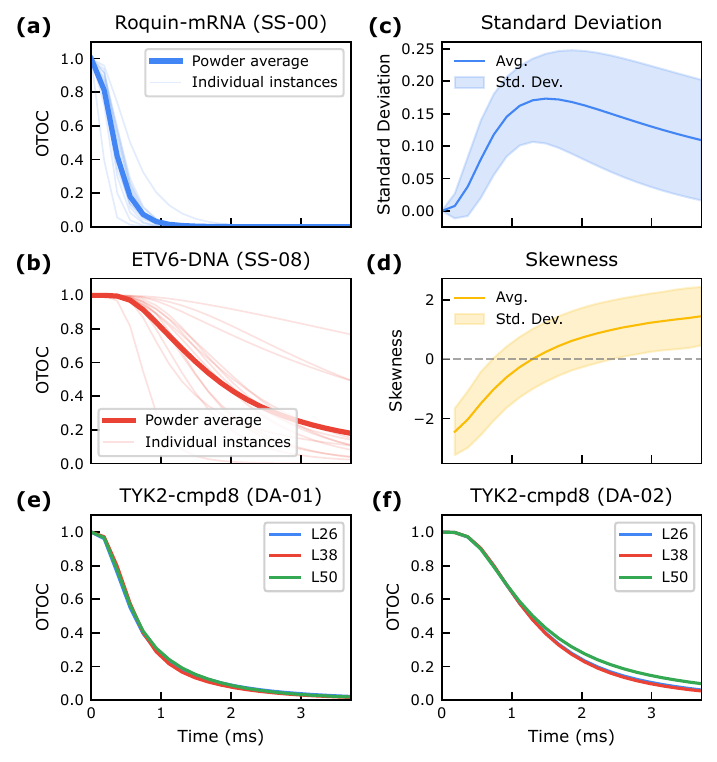}
    \caption{The powder averaged OTOC compared with individual instances. (a) An example of the powder average with carbons that are relatively close (5.3~\AA), and (b) an example with carbons that are further away (13.3~\AA). Some statistics showing (c) the width of the distribution over instances averaged over different systems vs time and (d) the skewness of the distribution of vs. time. In panels (e) and (f) we show how the powder average changes with increasing size of the Lebedev grid.}
    \label{fig:PA}
\end{figure}
In Fig.~\ref{fig:PA} we show how the powder-averaged OTOC compares to the individual instances for the 13 unique OTOCs in a 26-point Lebedev quadrature. There is a significant spread in the individual OTOCs, but, as shown in Fig.~\ref{fig:PA}(d), the distribution is not significantly skewed away from the mean. Near the endpoints of the OTOC there is a natural skew to the distribution due to the proximity of the average to the hard upper or lower limit, but in between there is very little skew. This suggests that the powder average should converge with relative ease as shown in Fig.~\ref{fig:PA}(e) and (f). That being said, the importance of the powder average is apparent in the large spread of the individual instances, which is far greater than the resolution we expect to need to meaningfully distinguish different structures.

\section{Results: Classical simulation}\label{sec:results_classical_sim}
The results in Sec.~\ref{sec:results_otocs} were all computed with exact statevector simulation (see Sec.~\ref{sec:results_exact_sim} below for details), but a natural question is whether some approximate simulation method is sufficient. In Sections~\ref{sec:Pauli_propagation}-\ref{sec:results_dtwa} we explore the accuracy of several types of approximate simulation methods applied specifically to the OTOCs studied in this work.

\subsection{Exact simulation}\label{sec:results_exact_sim}
Exact simulation was performed in all cases as described in Sec.~\ref{sec:pipeline} for $n = 14 - 22$ spin clusters. For dataset-C we simulated $n = 8 - 22$ so that the smaller spin systems could be used as benchmarks for approximate simulation methods. We used a first-order Trotter formula based on splitting the Hamiltonian into three mutually commuting groups of $XX$, $YY$, and $ZZ$ terms. The total time interval was split into 20 subintervals and each small interval was split into 15 Trotter steps. This results in a Trotter step size of $\approx 0.012$ ms ($\approx 0.006$ ms for the static structures). 20 random states were used to sample the trace at each time. See Appendix~\ref{sec:App_exact_sim_error} for further analysis of the convergence with respect to the number of Trotter steps and the sampling. All ``exact'' OTOCs reported in this work are computed using this procedure.

\subsection{Pauli propagation}
\label{sec:Pauli_propagation}
Pauli propagation and related methods~\cite{rall2019simulation,rudolph2023classical,beguvsic2024fast,angrisani2025classically,beguvsic2025real,rudolph2026pauli,beguvsic2025simulating} provide a systematic framework for approximating the trace or expectation value of a time-evolved operator. For the OTOCs considered here, our strategy was to compute an approximate Pauli representation of the time evolved butterfly operator
\begin{equation}
    X_b(t) \approx \sum_i c_i P_i.
\end{equation}
Here we have used $i$ to index all possible Pauli strings, $\{P_i\}$.
The OTOC can then be computed as 
\begin{align}
     \mathrm{OTOC} &= \frac{1}{2^n}\sum_{ij}\sum_{\mathbf{q}} c_ic_j\matrixel{\mathbf{q}}{
    P_iX_mP_jX_m}{\mathbf{q}} \\
    &= \frac{1}{2^n}\sum_{ij}\sum_{\mathbf{q}} c_ic_js_j\matrixel{\mathbf{q}}{
    P_iP_j}{\mathbf{q}} \\
    &= \sum_i c_i^2s_i\label{eqn:pauli_otoc}
\end{align}
where $s_i$ is a sign that depends on whether $X_m$ anticommutes ($s_i = -1$) or commutes ($s_i = 1$) with $P_i$. The heart of the approximation comes from how the Pauli representation is truncated. Pauli propagation is most efficient when high-weight Pauli strings can be discarded, but even when this is not the case, truncating the expansion based on the absolute size of the coefficients provides a systematically improvable approximation.

Pauli propagation was applied to all C-C pairs in dataset-C with truncation to a fixed pool size based on the magnitude of the coefficients. Other truncation schemes are possible, and we discuss some further results in Appendix~\ref{sec:App_pauli_extra}. In all calculations presented here we used a first order Trotter formula for the time evolution with the Hamiltonian terms grouped into three mutually commuting $XX$, $YY$, and $ZZ$ terms. Slightly larger trotter error was observed for coarse truncation, so a finer step-size was used: each $\approx 0.18$  ms time interval was divided into 60 Trotter steps. The Pauli-string representation of the operator is truncated based on the absolute value of the coefficients to maintain a fixed-sized ``pool'' of operators at each time step.

\begin{figure}
    \centering
    \includegraphics[scale=0.7]{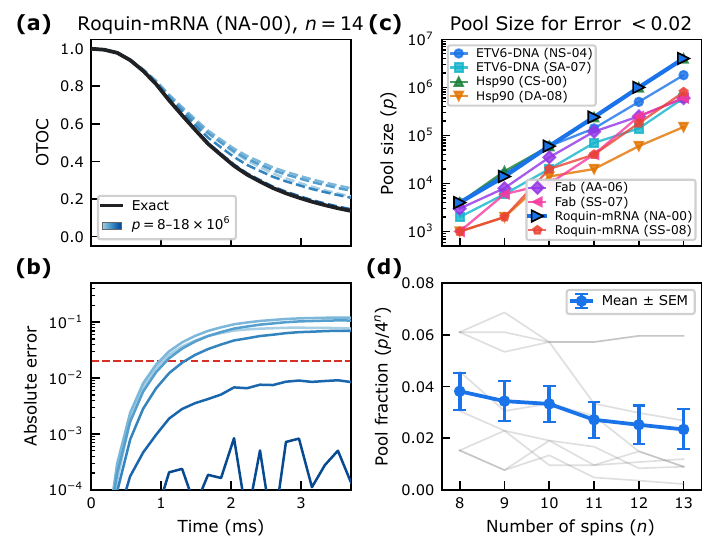}
    \caption{Results of Pauli propagation. (a) Approximate OTOCs and (b) error for an example, ($n = 14$ spins) at different pool sizes. (c) The necessary pool size and (d) the mean pool fraction to achieve a fixed error as a function of the spin network size ($n$) with errors bars showing the standard error in the mean (SEM).}
    \label{fig:pauli_prop}
\end{figure}

The results are shown in Fig.~\ref{fig:pauli_prop}. For the systems shown here, the cost to achieve a fixed error scales exponentially in the size of the spin network as we would expect from a naive truncation strategy. There is some indication (Fig.~\ref{fig:pauli_prop}(d)) that the scaling may be weaker than a single exponential at large $n$, but this is of little practical importance given the very large number of terms needed to attain high accuracy even at relatively modest spin cluster sizes. While this suggests that Pauli propagation based on standard truncation methods will fail at approximating the OTOC, it is still possible that a truncation scheme that uses the structure of the Hamiltonian more directly could be more efficient. 

\subsection{Truncated Liouvillian}
Truncated Liouville-space simulations provide a systematic approximation scheme for spin systems in which operator correlation growth remains sufficiently bounded~\cite{KUPROV2007241, BUTLER2009377}. In such cases, rather than keeping track of the state in its entirety, the dynamics are restricted to the subspace spanned by Pauli strings with weight below a prescribed cutoff ($k$). This uniform truncation is independent of the spatial support of the Pauli strings and does not exploit any internal structure in the operator correlations. Denoting the corresponding projector by $\mathcal P_k$, the projected Heisenberg equation of motion is given by
\begin{equation}
\frac{dX_b^{(k)}(t)}{dt}
= \mathcal P_k \mathcal L^{\dagger} \mathcal P_k X_b^{(k)}(t)
\equiv \mathcal L_k^{\dagger} X_b^{(k)}(t),
\label{projected_liouville}
\end{equation}
with $\mathcal{L}(\cdot)\equiv-i[H_{\rm eff}, \cdot]$.
As the projector $\mathcal P_k$ is Hermitian and the super operator $\mathcal L_k^{\dagger}$ is anti-Hermitian, the truncated propagator $e^{\mathcal L_k t}$ remains unitary on the truncated subspace. The norm of the retained operator content is therefore conserved, in contrast to the coefficient size truncation Pauli propagation scheme of Sec.\ref{sec:Pauli_propagation}, which sheds norm as small magnitude strings are removed.  
The resulting operator can be expanded as
\begin{equation}
X_b^{(k)}(t)
= \sum_{\{i|w(P_i)\leq k\}} c_i^{(k)}(t) P_i,
\end{equation}
where $w(P_i)$ denotes the Pauli weight of $P_i$, defined to be the number of non-identity operators in the string. In practice, the projected equation Eq.~(\ref{projected_liouville}) is computed numerically by combining $\mathcal{L}_k$ with standard time propagation schemes -- such as matrix exponentiation for small $k$, Krylov subspace methods~\cite{KUPROV200845}, or Trotter splitting methods for larger $k$. The coefficients $c_i^{(k)}(t)$ generated by the projected dynamics generally differ from those obtained by truncating the exact Heisenberg operator $X_b(t)$, but the approximation is expected to be accurate for local, weakly scrambling dynamics, for which operator spreading remains limited. Because our operator growth is seeded locally, low-weight truncations can also capture short-time dynamics in more strongly scrambling systems, although they need not reproduce the full OTOC time dynamics. Increasing $k$ systematically enlarges the retained operator space, and in the limit ($k\rightarrow n$), the full operator space is recovered and the dynamics become exact.

\begin{figure}
    \centering
    \includegraphics[scale=0.7]{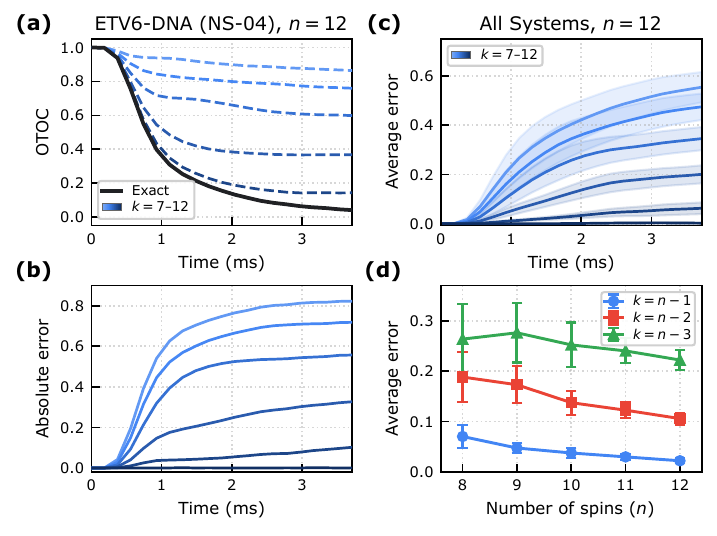}
    \caption{(a) Truncated Liouville-space OTOC simulations with varying truncation levels for an example system of 12 spins and (b) the error in the OTOC for the same system.  (c) The average OTOC error as a function of time for different $k$ averaged over all systems for $n = 12$ spins. (d) The average error vs cluster size for $k = n - 1, n - 2, n - 3$.}
    \label{fig:Truncated_Liouvillian}
\end{figure}
Truncated Liouville-space simulations were implemented in Spinach~\cite{hogben2011spinach}, utilizing the IK-0 basis set constructor and Krylov subspace propagation. The IK-0 basis set constructor allows a selective truncation corresponding to Pauli weight, $k$, which was varied from 2 to $n$, where $n$ is the total number of spins. The same calculations were implemented in a standalone Python code. The time propagation was implemented using the {\tt expm\_multiply} function from SciPy~\cite{virtanen2020scipy} with the Liouvillian stored as a sparse matrix. We verified that both the Spinach and Python implementations agree to high precision. The results in Fig.~\ref{fig:Truncated_Liouvillian} show powder averaged OTOC curves at increasing truncation level $k$ for one example system in dataset-C ($n=12$). Low weight truncations do not effectively capture the OTOC decay, and convergence to the exact result is only observed for the exact $k = n$ calculation. The convergence with $k$ is systematic, but even for $k=n-1$ we see a large deviation from the exact result. Fig.~\ref{fig:Truncated_Liouvillian}(c) shows the growth of the OTOC error for $k=7-12$, which mirrors the behavior of the single example shown in (a) and (b). Fig.~\ref{fig:Truncated_Liouvillian}(d) shows the average error as a function of the size of the spin cluster for $k = n - 1, n - 2, n -3$. The error is nearly constant with respect to the spin cluster size, suggesting that the method will scale nearly exponentially with $n$. In practice, system sizes that are accessible with exact simulation require $k \approx n$ for an acceptable error.

\subsection{Matrix product operator simulation}
Just as a wavefunction can always be exactly decomposed as a matrix product state (MPS), a general operator in the many-body space can be similarly decomposed into a matrix product operator (MPO):
\begin{align}
    O &= \sum_{\mathbf{z}}O^{z_1z_2\ldots z_{n}}o_{z_1}o_{z_2}\ldots o_{z_n} \\
    &= \sum_{\mathbf{z}}\mathbf{W}_1^{z_1} \mathbf{W}_2^{z_2} \ldots \mathbf{W}_n^{z_n} o_{z_1}o_{z_2}\ldots o_{z_n}.
\end{align}
Here we have used $z_i$ to index some complete set of operators acting on a single site, $\{o_{z_i}\}$. MPOs are commonly used in implementations of the density matrix renormalization group and related MPS methods\cite{verstraete2004matrix,mcculloch2007density}, and they have been used effectively to simulate OTOCs with local, 1-dimensional Hamiltonians\cite{hemery2019matrix,xu2020accessing,lopez2021operator}. Our approach here is similar to what was used in the supplementary information of Ref.~\cite{zhang2025quantum} (differing from Refs.~\cite{hemery2019matrix,xu2020accessing,lopez2021operator} only in that we must deal with the all-to-all connectivity of the Hamiltonian). To compute the OTOC, we first compute an approximate MPO representation of the time-evolved butterfly operator, $\tilde{X}_b(t)$, from which we can easily form an MPO representation of $\tilde{X}_b(t)X_m$. The approximate OTOC,
\begin{equation}
    \frac{1}{2^n}\mathrm{Tr}\left[(\tilde{X}_b(t)X_m)^2 \right],
\end{equation}
can then be computed by contracting the MPO, $\tilde{X}_b(t)X_m$, with itself. We perform the time evolution by grouping the pairwise terms in the Hamiltonian into commuting groups and then implementing the Trotterized time evolution. Each term in the Trotterized time evolution acts on a pair of sites, and it is implemented by swapping the locations of the sites in the MPO until the two sites are adjacent and then exactly applying the local two-site unitary.

This method is approximate because the maximum bond dimension of the MPO is always truncated to a specified value, $\chi$. Every time two sites are swapped, the bond dimension, which can grow to implement the swap, is truncated back to a maximum value of $\chi$, and the same truncation is applied after each local gate. In practice, the truncation error is dominated by swapping sites even though grouping the terms into disjoint (commuting) groups minimizes the number of swaps that is needed for each group. We also try to reduce the truncation error by reordering the sites at the beginning of the computation using the Fiedler ordering based on the Hamiltonian matrix elements\cite{olivares2015ab}. This entire procedure is implemented, largely using built-in methods, with quimb\cite{gray2018quimb}.

\begin{figure}
    \centering
    \includegraphics[scale=0.7]{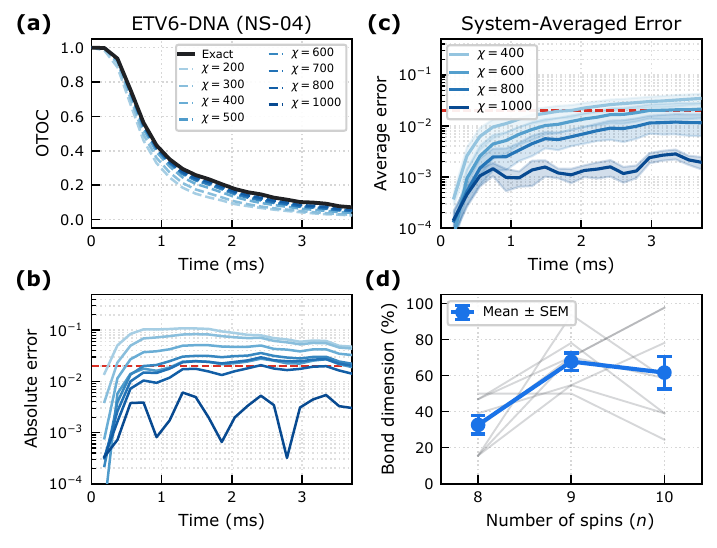}
    \caption{(a) OTOCs from MPO simulation with different max. bond dimension ($\chi$) and (b) the error in the OTOC at different $\chi$ in an example system (ETV6-DNA, NS, 04, $n = 10$ spins). (c) The error in the OTOC for different bond dimensions averaged over all 8 examples in dataset-C. (d) The bond dimension needed to achieve OTOC error $< 0.02$ as a function of the number of spins. The bond dimension is plotted as a percentage of the maximum bond dimension ($\chi_{\rm{max}} = 4^{\lfloor n/2 \rfloor })$.}
    \label{fig:mpo}
\end{figure}

MPO simulations were performed on dataset-C for $n = 8 - 10$ spins. All calculations were performed with the same number of Trotter steps as the exact calculation (15 Trotter steps for each $\approx 0.18$ ms time interval). The results are shown in Fig.~\ref{fig:mpo}.
The MPO OTOCs converge to the exact curve, but a large bond dimension is required to get an acceptable error (OTOC error $< 0.02$). For $n = 10$ the max bond dimension is 1024, and, as shown in Fig.~\ref{fig:mpo}(c) and (d), a significant fraction of the max bond dimension is required to get accurate results. The cost of the MPO simulation is expected to scale exponentially with the number of spins for 3D systems. While there could still be a practical speedup relative to exact simulation for modest system sizes, that is not what we observe in this case: we estimate that the $n = 12$ MPO simulations are already more expensive than the exact $n = 22$ simulations. Note that we have not thoroughly optimized the implementation, and we have not tried additional time propagation schemes, so there may be considerable room for improvement. In particular, propagation methods based on the time-dependent variational principle (see for example Ref.~\cite{hemery2019matrix}) may be significantly more efficient for these 3D systems with long range interactions. 

\subsection{Discrete truncated Wigner approximation}\label{sec:results_dtwa}
We also consider a semiclassical approximation called the discrete truncated Wigner approximation (DTWA)\cite{wootters1987wigner,schachenmayer2015many,shirai2025out}. To motivate this approximation, we first define the phase point operator
\begin{equation}
    A(\mathbf{s}_{\gv{\tau}}) \equiv \prod_{k} \frac{1}{2}\left[I + \mathbf{s}_{\tau_k}\cdot \gv{\sigma}_k \right],
\end{equation}
where $\mathbf{s}_{\gv{\tau}}$ is a tuple of spin direction vectors for each ($k \in 1, \ldots, n$) spin and $\gv{\sigma}$ is the vector of Paulis. We choose to work in the discrete product space spanned by products of
\begin{align}
    \mathbf{s}_{(0,0)} &\equiv (1, 1, 1) \\
    \mathbf{s}_{(0,1)} &\equiv (-1, -1, 1) \\
    \mathbf{s}_{(1,0)} &\equiv (1, -1, -1) \\
    \mathbf{s}_{(1,1)} &\equiv (-1, 1, -1).
\end{align}
Given these definitions, the Wigner function for a particular phase-space point is given by
\begin{equation}
    W_{\gv{\tau}} = \frac{1}{2^n}\mathrm{Tr}\left[\rho A(\mathbf{s}_{\gv{\tau}})\right],
\end{equation}
such that any time-evolved observable may be computed as
\begin{align}
    \avg{O(t)} &= \sum_{\gv{\tau}} W_{\gv{\tau}}\mathrm{Tr}\left[A(\mathbf{s}_{\gv{\tau}}) e^{iHt}Oe^{-iHt}\right] \nonumber \\
    &= \sum_{\gv{\tau}} W_{\gv{\tau}}\mathrm{Tr}\left[e^{-iHt}A(\mathbf{s}_{\gv{\tau}}) e^{iHt}O\right].
\end{align}
We now make the semiclassical truncated Wigner approximation:
\begin{equation}
    e^{-iHt}A(\mathbf{s}_{\gv{\tau}})e^{iHt} \approx A(\mathbf{s}_{\gv{\tau}}(t)).
\end{equation}
Instead of propagating the phase-point operator backwards in time, an exact approach, we approximate it as the operator evaluated with classical spin vectors that have been propagated forward in time using a classical equation of motion:
\begin{equation}
    \frac{d}{dt}\mathbf{s}_{\tau_k}(t) = -2 \mathbf{s}_{\tau_k}(t)\times \frac{\partial H}{\partial \mathbf{s}_{\tau_k}}.
\end{equation}
The application of this method to evaluating OTOCs is adapted from Ref.~\cite{shirai2025out} and described in more detail in Appendix~\ref{sec:App_dtwa}. Unlike the other classical simulation methods we consider, the DTWA is not systematically improvable in general. However, there are perturbative approaches that can be used to go beyond the DTWA\cite{polkovnikov2010phase,schachenmayer2015many}, and recently introduced cluster-based extensions to the method\cite{braemer2024cluster,nagao2025two} are theoretically systematic with increasing cluster size.

\begin{figure}
    \centering
    \includegraphics[scale=0.7]{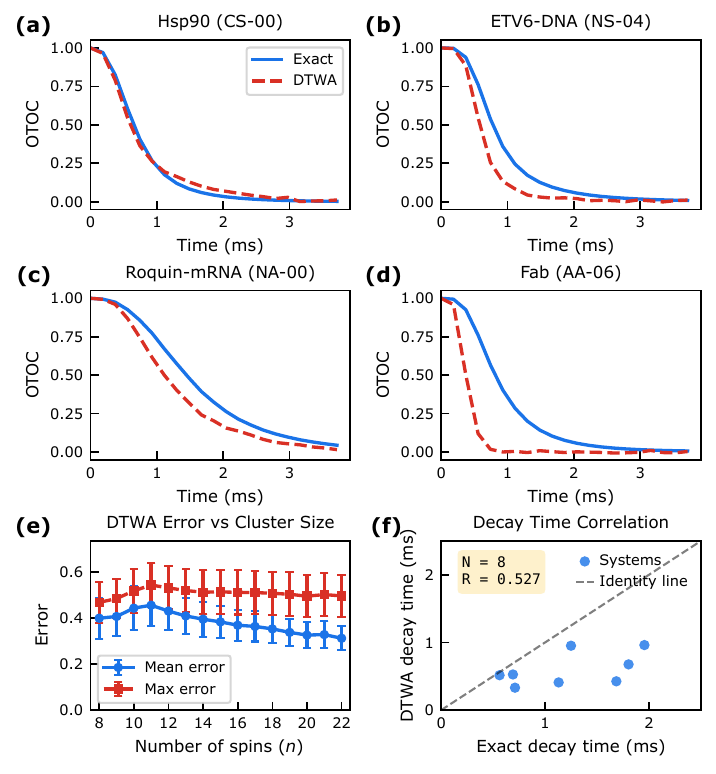}
    \caption{The DTWA applied to OTOCs in protein NMR. (a) - (d) Examples of comparing the exact OTOC to the DTWA OTOC are shown for $n = 22$. (e) The average and max error is shown as a function of the number of spins with error bars showing the standard error in that quantity over the 8 examples. For each example, the error is averaged only over the points where the exact OTOC falls between 0.8 and 0.2. so that we avoid an artificially low error from larger plateau regions. (f) The correlation between DTWA and exact OTOC decay times (the time for the OTOC to decay to 0.6).}
    \label{fig:dtwa}
\end{figure}
The DTWA was applied to all 8 C-C pairs in dataset-C, and in each case, we used a time step of $\approx 0.007$ ms with a 4th order Runge-Kutta integrator and 1000 trajectories for each instance in the powder average. The results are shown in Fig.~\ref{fig:dtwa}. While the DTWA can easily scale to hundreds of spins and performs very well in some cases, the error is too large for it to be generally useful at approximating OTOCs for the systems considered here. However, there is reason to believe that the underlying semiclassical approximation may be less severe at larger system sizes, and we do observe that the average error decreases slightly with increasing cluster size. This is somewhat consistent with DTWA performing better for higher dimensional systems with longer-range interactions\cite{schachenmayer2015many}. The DTWA systematically underestimates the OTOC decay time (see Fig.~\ref{fig:dtwa}(f)), but there is still some correlation with the exact OTOC decay times. Despite the relatively poor accuracy for $n = 8 - 22$ shown here, semiclassical methods like the DTWA may still be useful in practice if the correlation with the exact OTOC improves for larger $n$.

\section{Discussion}\label{sec:discussion}
We have presented a framework for modeling OTOCs in protein systems under conditions relevant to realistic solid-state NMR experiments. As a demonstration of this framework, we have provided a dataset of over 350 OTOCs in 10 protein systems. In the short term, our aim is to provide realistic examples for study by the community, and in this work we have already studied several aspects of these OTOCs over the entire dataset. In the longer term, we hope that this is a step towards simulating OTOCs that are also implemented in real NMR experiments for protein structure determination.

The OTOCs studied here differ from idealized models in a number of ways: (1) we use real protein structures and select carbon-carbon pairs that are representative of distances that would be useful for structure determination; (2) we use an effective Hamiltonian that is realizable in experiments; (3) we include both averaging over fast molecular motion and powder averaging. Accounting for all these details makes these OTOCs qualitatively representative of what we would like to experimentally implement, but many details will depend on a particular experimental setup. Some other details, like higher order terms in the effective Hamiltonian, are not included here and they may need to be accounted for to match an experiment.

\subsection{Information propagation and implications for NMR}
One of the main takeaways from this work, as shown in Fig.~\ref{fig:operator_front_all}, is that we can extract system-dependent operator spreading velocities from linear fits across almost all the systems studied here. The fact that the velocities appear approximately constant is interesting because it suggests that the effective dimensionality of these systems is closer to 1 over short distances. Furthermore, the propagation velocities allow us to estimate characteristic OTOC evolution times of 2.67–10.29 ms, depending on the spin system. Although these estimates are based on effective Hamiltonian simulations, they indicate that OTOC dynamics occur on experimentally accessible timescales, comparable to those routinely employed in solid-state NMR experiments~\cite{Nishiyama_UFMAS_2023}.

To put these timescales into perspective, conventional $^{13}$C-$^{13}$C distance measurements based on effective double-quantum Hamiltonians solely rely on direct dipolar interactions between the carbon spins. The results in Ref.~\cite{PILEIO200765} show that even after powder averaging, the characteristic timescales of the dynamics remain governed by the effective dipolar coupling, decreasing with the inverse cube of the internuclear distance. For example, two $^{13}$C spins separated by 6~\AA~have a direct dipolar coupling of approximately 35 Hz, corresponding to an inverse-coupling timescale of approximately 30 ms, even before accounting for potential scaling factors. This comparison suggests that OTOCs may provide access to spatially extended spin correlations on considerably shorter timescales, which could be particularly advantageous for probing larger spatial separations.

We also have shown evidence that only a single structural parameter can be extracted from one OTOC. This is important in considering the situations where this technique can be most impactful for structure determination (see Sec.~\ref{sec:model_fitting_structure}). In a model problem, this single structural parameter was largely composed of just the interspin distance. If this holds in general, then the learning problem may be just a matter of fitting one distance per OTOC. However, the system-dependence of the operator-spreading velocities would suggest that this fitting has to be performed separately for each system by comparing to an accurately simulated OTOC.

\subsection{Implications for classical simulation}
Given that accurate simulation may be an important component of learning structure from OTOCs, we have tried a number of common approaches. None of the approaches we tried were able to achieve high accuracy at a reasonable cost, but the set of approaches we tried is far from exhaustive. For example, a more tailored approach, like Pauli propagation with a custom truncation scheme or MPO simulation with a custom time-propagation method, could perform better. We also have not tried any diagrammatic methods. The selection of methods shown here is enough to suggest that classical simulation for larger system sizes is non-trivial, but it is far from the end of the story. Classical simulation of realistic OTOCs is an important future direction since quantum simulation will be the only avenue left if no viable classical technique is found.

\subsection{Model fitting and structure determination}\label{sec:model_fitting_structure}
NMR is most commonly used as an analytic tool; the information encoded in the nuclear spin dynamics can be used to determine molecular structure.
In this sense, the pipeline in Sec.~\ref{sec:pipeline} describes a method of solving the problem $\mathrm{Model}\rightarrow \mathrm{OTOC}$, but in practice we desire to solve the inverse problem $\{\mathrm{OTOC}\}\rightarrow\mathrm{Model}$ after measuring a dataset $\{\mathrm{OTOC}\}$ experimentally.
This presents two challenges: it is not always possible to solve the inverse problem from the amount of data given, and even when this is possible it can be a hard search problem to find the correct model.

To solve the inverse problem at all, we need the measured data to be sufficiently sensitive to our model parameters.
We quantified this in Sec.~\ref{sec:results_information}, by looking at the eigenvalues of the pullback metric.
Small eigenvalues of this metric correspond to linear combinations of parameters that do not change the predictions of the model upon being shifted.
If the number of non-negligible eigenvalues is significantly less than the number of model parameters, the inverse problem cannot be solved with the data given.
More generally, the effective dimension $D_{\mathrm{eff}}$ roughly quantifies the number of parameters that can be learned from the given data; for a single OTOC this appears to be around $1$--$2$.
This suggests a simple rule of thumb: OTOCs between $M$ distinct butterfly--measurement pairs can constrain roughly $M$ parameters.
Confirming this in a larger system remains an outstanding task.

However, specifying even a static molecule generally requires a number of parameters that scales linearly with the number of atoms, $\mathcal{O}(N)$. Thus, if $M\ll 3N-6$, it can be advantageous to reparameterize the system in terms of a smaller set of collective variables. Reparameterization also shrinks the search space for an optimal model, which can be necessary to avoid worst-case exponential scaling (even with help from a quantum computer).\footnote{The search problem is also simplified by access to a good initial guess, or if individual datapoints are each sensitive to only a few parameters.}
For example, the dimethylbiphenyl system of Ref.~\cite{zhang2025quantum} was reparameterized in terms of the dihedral angle, with other parameters fixed via a combination of MD and force-field fitting.
The dihedral angle was chosen as chemical intuition suggests classical methods struggle to balance the competing forces acting on this degree of freedom.

If chemical intuition does not scale to a larger molecule, we suggest similar collective variables can be obtained from MD itself. 
Standard methods such as principal component analysis~\cite{amadei1993essential,palma2023uses}, timescale analysis~\cite{naritomi2011slow,naritomi2013slow}, and machine-learning-based approaches~\cite{sultan2018automated,bonati2020data} are routinely used to construct collective variables from MD simulations.
It is often the case that the most interesting modes are slow, floppy modes with a wide distribution.
These modes are also the most difficult to model with MD due to the long timescales needed for good sampling and are therefore good candidates for experimental determination.
Furthermore, these collective-variable frameworks can easily incorporate constraints from other experimental techniques~\cite{granata2013characterization,brotzakis2019enhanced} or from AlphaFold structures~\cite{spiwok2022collective,vani2023alphafold2}.
We are hopeful that, in combination with other experimental methods, a few judiciously chosen OTOCs can significantly improve the effective resolution for precisely those degrees of freedom that are most biologically interesting.

\subsection{Open problems}
The work presented here is just the beginning of what is necessary to combine chemical modeling, classical or quantum simulation, and NMR experiments into a useful framework for structure determination. In particular, here are some important open problems suggested by this work:
\begin{itemize}
    \item {\bf Testing and designing classical simulation methods for OTOCs in NMR}: The set of classical simulation methods investigated between this work and Ref.~\cite{zhang2025quantum} is still not complete, nor has work been done yet to design classical methods specifically to simulate OTOCs in Hamiltonian systems.
    \item {\bf Implementing model fitting}: Model fitting is the key step for structure determination, and this still needs to be implemented and tested for realistic problems.
    \item {\bf Efficient simulation of the actual sequence}: In actual experiments, the time-dependence of the real pulse sequence and decoherence may require a more complex simulation that includes this time-dependence along with open system dynamics. These more detailed simulations may be necessary to compare with experiment and to determine if the conclusions of this work will extend to an actual experiment.
    \item {\bf Finding optimal sequences}: Finding optimal pulse sequences to implement the target effective Hamiltonian with time reversal will be necessary for experimental realization of the OTOC in protein systems.
    \item {\bf Improving quantum simulation techniques for OTOCs}: Optimized quantum algorithms for simulating OTOCs, such as improved product formula methods, still need to be developed. This is potentially important both for pushing the limits of currently available devices and for guiding hardware development.
    \item {\bf Strongly-correlated dynamics beyond OTOCs}: In this work we have focused on a study of OTOC(1), with a first look at the structural sensitivity of OTOC(2) in the proxy model of Sec.~\ref{sec:results_information}. Higher order~\cite{google25constructive} or multi-dimensional extensions may have different advantages in timescale, sensitivity, and experimental design, which remain to be explored in realistic protein systems.
\end{itemize}
To summarize, we have taken the first steps toward developing the theory and modeling necessary to enable OTOC-NMR as a viable tool for protein structure determination, but there is much more left to do.

\section*{Acknowledgments}

The authors thank Ashok Ajoy, Chandrasekhar Ramanathan, and Ryan Babbush for fruitful discussions on this topic.

\section*{Code and data availability}
All protein structures, effective Hamiltonians, and exact and approximate OTOCs will be made publicly available along with the code used to generate the effective Hamiltonians and exact OTOC curves.

\section*{AI use declaration}

AI was used to assist code generation (Gemini 3.1 Pro and Gemini 3.5 Flash), and for proofreading and polishing purposes (Gemini 4 Argon and Gemini 4 Carbon).
All code was verified by extensive unit testing and checking by humans, and the authors take responsibility for all data produced and all conclusions reached in the work.

\clearpage

\appendix

%%%%%%%%%%%%%%%%%%%%%%%%%%%%%%%%%%%%%%%%%%%
\section{NMR Primer for QIS}\label{app:NMR_primer}
%%%%%%%%%%%%%%%%%%%%%%%%%%%%%%%%%%%%%%%%%%%

%%%%%%%%%%%%%%%%%%%%%%%%%%%%%%%%%%%%%%%%%%%
\subsection{NMR Notation and Basics}
%%%%%%%%%%%%%%%%%%%%%%%%%%%%%%%%%%%%%%%%%%%
We first establish the correspondence between the notation used in quantum information science (QIS) and that conventionally used in NMR. For spin-1/2 nuclei, the NMR spin operators are related to the Pauli operators by
\begin{align}
    I_{ix} = \frac{1}{2}X_i, \quad I_{iy} &= \frac{1}{2}Y_i, \quad I_{iz} = \frac{1}{2}Z_i, \\
    \vec{I}_i &= \frac{1}{2}\vec{\sigma}_i,
\end{align}
where $X_{i}, Y_{i}, Z_{i}$ are the Pauli operators acting on spin $i$. Apart from notational differences, for spin-1/2 nuclei, the NMR and QIS operator conventions thus differ only by a normalization factor. When multiple nuclear species are present, NMR conventionally distinguishes between them by denoting their spin operators as $\vec{I}$ and $\vec{S}$, respectively. This is typically not done in QIS, where the distinction is left to the qubit indexing that adopts a linear scheme. For clarity, for a spin system of $N_I$ spins of one species, and $N_S$ spins of another, the exact translation from NMR spin operators to QIS Pauli operators goes as: 
\begin{align}
    &\{\vec{I}_1,\ldots,\vec{I}_{N_I}, \vec{S}_{1},\ldots, \vec{S}_{N_S}\} \nonumber \\ &\rightarrow \frac{1}{2}\{\vec{\sigma}_1,\ldots,\vec{\sigma}_{N_I}, \vec{\sigma}_{N_I+1},\ldots, \vec{\sigma}_{N_I+ N_S}\}.
\end{align}
We use NMR notation when describing the underlying NMR interactions, while expressing the final results in the QIS Pauli-operator notation where appropriate.

The NMR spin Hamiltonian can be decomposed into contributions from the different interaction terms
\begin{equation}
H=\sum_\Lambda H^\Lambda,
\end{equation}
where $\Lambda$ labels the different types of interactions. A general interaction term can be written in the Cartesian representation as 
\begin{equation}
    H^\Lambda = C^\Lambda (\vec{A}^\Lambda)^\mathrm{T} \cdot W^\Lambda \cdot  \vec{B}^\Lambda,
\end{equation}
where $C^\Lambda$ is a characteristic interaction scalar, $W^\Lambda$ is a rank-2 Cartesian coupling tensor and $\vec{A}$ and $\vec{B}$ are vector operators of the two interacting quantities. For spin–spin interactions, these quantities are nuclear spin operators; for spin–field interactions, one quantity is a spin operator and the other is an applied magnetic field.

The total magnetic field can be written as
\begin{equation}
\begin{gathered}
\vec{\mathrm{B}}(t)=\vec{\mathrm{B}}_{0}+\vec{\mathrm{B}}_{\rm rf}(t),
\end{gathered}
\end{equation}
where $\vec{\mathrm{B}}_{0}$ is the static Zeeman field and $\vec{\mathrm{B}}_{\rm rf}(t)$ describes the application of time-dependent applied radio-frequency fields. The static magnetic field is typically applied along the laboratory $z$-axis
\begin{equation}
\begin{gathered}
\vec{\mathrm{B}}_{0}=B_{0}\hat{e}_{z}.
\end{gathered}
\end{equation}
The applied radio-frequency field is typically a linearly oscillating magnetic field oriented along the laboratory $x$-axis
\begin{align}
\vec{\mathrm{B}}_{\rm rf}(t)&=B^{I}_{1}(t)\cos(\omega_{I}t+\phi_{I}(t))\hat{e}_{x}\nonumber \\ &+B^{S}_{1}(t)\cos(\omega_{S}t+\phi_{S}(t))\hat{e}_{x},
\end{align}
where $\omega_{I}$ and $\omega_{S}$ are the RF frequencies applied near the respective resonance frequencies of the $I$ and $S$ spins. In the resonant case, these frequencies are matched to the Larmor frequencies $\omega^{0}_{I}=-\gamma_{I}B_{0}$ and $\omega^{0}_{S}=-\gamma_{S}B_{0}$. In the rotating-frame approximation, the radio-frequency and Zeeman Hamiltonian reduce to
\begin{align}
&H^{\rm Z}_{\rm rot}=\sum_i(\omega^{0}_{I}-\omega_{I})I_{iz}
+\sum_m(\omega^{0}_{S}-\omega_{S})S_{mz},
\\
&H^{\rm rf}_{\rm rot}=-\sum_i \frac{\gamma_{I}}{2}B^{I}_{1}(t)(\cos(\phi_{I}(t))I_{ix}+\sin(\phi_{I}(t))I_{iy})\nonumber \\ &- \sum_m\frac{\gamma_{S}}{2}B^{S}_{1}(t)(\cos(\phi_{S}(t))S_{mx}+\sin(\phi_{S}(t))S_{my}).
\end{align}

The magnetic fields described above, together with the spin–spin interactions, determine the NMR spin Hamiltonian. Although several terms may contribute to the dynamics, we are specifically interested in the dipolar contribution. As discussed below, pulse sequences may be designed to selectively isolate this contribution from the other terms in the effective Hamiltonian. We therefore restrict the following discussion to $H^{\Lambda}=H^{\mathrm{D}}$, which describes the effective dynamics of interest. For two nuclear spin species, it is convenient to express the dipolar Hamiltonian in terms of the corresponding NMR spin operators, $\vec{I}$ and $\vec{S}$
\begin{align}
H^{\mathrm{D}}=&
\sum_{i<j}
C^{\mathrm{D}}_{ij}
(\vec{I}_{i})^{\mathrm{T}}
W^{\mathrm{D}}_{ij}
\vec{I}_{j}
+
\sum_{m<n}
C^{\mathrm{D}}_{mn}
(\vec{S}_{m})^{\mathrm{T}}
W^{\mathrm{D}}_{mn}
\vec{S}_{n}
\nonumber \\&+
\sum_{i,m}
C^{\mathrm{D}}_{im}
(\vec{I}_{i})^{\mathrm{T}}
W^{\mathrm{D}}_{im}
\vec{S}_{m}.
\label{eq:Dipolar_cartesian}
\end{align}
The first two terms correspond to homonuclear dipolar interactions, involving $I$–$I$ and $S$–$S$ spin pairs, respectively, while the final term describes heteronuclear dipolar interactions between $I$ and $S$ spins. For a pair of spins separated in space by the displacement vector $\vec{r}_{ij}$, the dipolar coupling tensor is
\begin{equation}
W^{\mathrm{D}}_{ij}=
3\hat{r}_{ij}\otimes \hat{r}_{ij}^{\mathrm{T}}-\mathbbm{1},
\end{equation}
where $\hat{r}_{ij}=\vec{r}_{ij}/\vert\vert\vec{r}_{ij}\vert\vert$ is the unit vector connecting the two spins. The corresponding interaction strength is
\begin{equation}
C^{\mathrm{D}}_{ij}=
-\frac{\mu_0}{4\pi}
\frac{\gamma_i\gamma_j\hbar}{r_{ij}^3} \equiv 2\pi b_{ij},
\end{equation}
where $\gamma_i$ and $\gamma_j$ are the gyromagnetic ratios of the two nuclei and $r_{ij}=\vert\vert\vec{r}_{ij}\vert\vert$ is their distance, and we also make explicit the relation to the dipolar coupling constants Eq.~(\ref{eq:dipolar_coupling_constants1}-\ref{eq:dipolar_coupling_constants3}), defined in the main text.

Since the spin operators are defined with respect to the laboratory-frame axes, we initially express the displacement vectors ($\vec{r}_{ij}$) in the same coordinate system. When working with powdered samples, however, the molecules are randomly oriented with respect to the laboratory frame. It is therefore more convenient to define the displacement vectors in a molecule-fixed frame and to describe its orientation relative to the laboratory frame separately
\begin{equation}
\vec{r}^{\rm L}_{ij}=R_{zyz}(\Omega_{\rm LM})\vec{r}^{\rm M}_{ij},
\end{equation}
where $R_{zyz}(\Omega_{\rm LM} = \Omega_{\mathrm{ML}}^{-1})$ is the Euler rotation relating the two frames. Since the dipolar interaction depends quadratically on the components of the internuclear displacement vector in the molecular frame, as well as on its orientation relative to the laboratory frame, its spatial dependence can be represented by a rank-2 spherical tensor,
\begin{equation}
    [A^{2q}_{ij}]^{\rm L} = \sum_{p=-2}^{+2}[A^{2p}_{ij}]^{\rm M}D^2_{pq}(\Omega_{\rm ML}),
\end{equation}
where $D^2_{pq}$ are Wigner D-matrix elements for the rank-2 irreducible representation of SO(3).

Irreducible spherical spin tensors are also defined, such that the dipolar Hamiltonian (\ref{eq:Dipolar_cartesian}) can be equivalently written 
\begin{align}
H^{\mathrm{D}}=&\sum_{q=-2}^{+2}(-1)^{q}\bigg\{
\sum_{i<j}
C^{\mathrm{D}}_{ij}
[A^{2q}_{ij}]^{\rm L}\mathcal{I}^{2-q}_{ij} 
\\&+
\sum_{m<n}
C^{\mathrm{D}}_{mn}
[A^{2q}_{mn}]^{\rm L}\mathcal{S}^{2-q}_{mn}
+
\sum_{i,m}
C^{\mathrm{D}}_{im}
[A^{2q}_{im}]^{\rm L}\mathcal{T}^{2-q}_{im}\bigg\}. \nonumber
\end{align}
Explicit expressions for the irreducible spin components are given in Table~\ref{table:spherical_tensors}. In the rotating-frame, and high magnetic-field approximation, the dipolar Hamiltonian further reduces to
\begin{align}
H^{\mathrm{D}}_{\rm rot}=&
\sum_{i<j}
C^{\mathrm{D}}_{ij}
[A^{20}_{ij}]^{\rm L}\mathcal{I}^{20}_{ij}
+
\sum_{m<n}
C^{\mathrm{D}}_{mn}
[A^{20}_{mn}]^{\rm L}\mathcal{S}^{20}_{mn} \nonumber
\\&+
\sqrt{\frac{2}{3}}\sum_{i,m}
C^{\mathrm{D}}_{im}
[A^{20}_{im}]^{\rm L}\mathcal{I}^{10}_{i}\mathcal{S}^{10}_{m}, 
\label{eq:secular_dipolar_Hamiltonian}
\end{align}
so that only the rotationally invariant part with respect to $z$ remains. Heteronuclear interactions are further truncated due to a substantial Larmor frequency difference between the $I$ and $S$ spins. 
The form of the rank-2 irreducible spatial components in the laboratory frame depends on if the sample is static, or mechanically spinning, discussed separately below. However, for a spin pair $X$ whose internuclear unit vector $\hat{\mathbf n}_X$ has polar and azimuthal angles $(\beta_X,\gamma_X)$ in a given frame, F, the dipolar spatial tensor is axially symmetric about
$\hat{\mathbf n}_X$ and its irreducible components in the frame F are
\begin{equation}
  [A^{2p}_X]^{\rm F} = \sqrt{6} d^{2}_{p0}(\beta_X)e^{\,ip\gamma_X}.
  \label{eq:spatial_components}
\end{equation}
Equivalently, we can write this in terms of the Cartesian components of $\hat{\mathbf n}_X$ in frame F, such that 
\begin{align}
  [A^{2\pm2}_X]^{\rm F} &= \tfrac{3}{2}\left([\hat{\mathbf n}_X]_x
                    \pm i[\hat{\mathbf n}_X]_y\right)^2 \\
  [A^{2\pm1}_X]^{\rm F} &= \mp 3\left([\hat{\mathbf n}_X]_x
                    \pm i[\hat{\mathbf n}_X]_y\right)[\hat{\mathbf n}_X]_z \\
  [A^{20}_X]^{\rm F}    &= \sqrt{\tfrac{3}{2}}
                   \left(2[\hat{\mathbf n}_X]_z^{2}
                        -[\hat{\mathbf n}_X]_x^{2}
                        -[\hat{\mathbf n}_X]_y^{2}\right),
\end{align}
which shows in principle how these components can be computed from a molecular geometry.

\subsubsection{Static Samples}
For the results presented in the main text, we assume a static powdered sample. In this case the chain of frame transformations for the spatial irreducible components is simply from the molecular frame to the laboratory frame. The dipolar Hamiltonian becomes

\begin{align}
H^{\mathrm{D}}_{\rm rot, static}=&\sum_{p=-2}^{+2}\bigg\{
\sum_{i<j}
C^{\mathrm{D}}_{ij}
[A^{2p}_{ij}]^{\rm M}D^{2}_{p0}(\Omega_{\rm ML})\mathcal{I}^{20}_{ij}
\nonumber \\
&+
\sum_{m<n}
C^{\mathrm{D}}_{mn}
[A^{2p}_{mn}]^{\rm M}D^{2}_{p0}(\Omega_{\rm ML})\mathcal{S}^{20}_{mn}
 \\
&+ 
\sqrt{\frac{2}{3}}\sum_{i,m}
C^{\mathrm{D}}_{im}
[A^{2p}_{im}]^{\rm M}D^{2}_{p0}(\Omega_{\rm ML})\mathcal{I}^{10}_{i}\mathcal{S}^{10}_{m}\bigg\}. \nonumber
\label{eq:static_dipolar_Hamiltonian}
\end{align}
 For a powdered sample, $\Omega_{\rm ML}$ are random variables and are, in practice, sampled from the powder averaging scheme. We may make the further identification with Eq.~\eqref{dipolar_coupling}, through Eq.~(\ref{eq:spatial_components}),
\begin{equation}
2\pi \sqrt{6}\tilde{b}_{ij}=C^{\rm D}_{ij}[A^{20}_{ij}]^{\rm L}=C^{\rm D}_{ij}\sum_{p=-2}^{+2}[A^{2p}_{ij}]^{\rm M}D^{2}_{p0}(\Omega_{\rm ML}),
\end{equation} 
to highlight the dipolar coupling constants and their explicit dependence on molecular orientation and distance.

\subsubsection{Spinning Samples}
In practice, solid-state NMR experiments are often performed under magic-angle spinning (MAS). This works by mechanically rotating the sample at a frequency $\omega_r$ about an axis inclined at the magic angle, $\beta_{\mathrm{ma}} = \arctan(\sqrt{2})$, to the external static magnetic field, $\vec{\mathrm{B}}_0$. Whilst having the advantage of improving spectral resolution due to rotational averaging, MAS also removes the dipolar
couplings that generate the OTOC dynamics, and therefore structural information, we wish to probe. To recover dipolar dynamics, one may utilize rotor-synchronized radio-frequency pulse sequences to selectively reinstate specific interaction components~\cite{doi:https://doi.org/10.1002/9780470034590.emrstm0551}. This is known as recoupling and is the result of breaking the rotational averaging that removes the interactions over a certain time interval. The resulting effective Hamiltonians are discussed further in
Appendix~\ref{sec:mas-design}.

To describe the Hamiltonian of a spinning sample, the rotor frame is introduced as an extra step in the transformation from the molecular to the laboratory frame. The Euler angles relating the rotor frame to the laboratory frame are $\Omega_{\rm RL} = (\alpha_{\rm RL}^0 - \omega_r t, \beta_{\rm ma},0)$, with $\alpha^0_{\rm RL}$ defining the initial position of the rotor. The two spin spatial spherical tensors in the laboratory frame become 
\begin{align}
    [A^{20}_{ij}]^{\rm L} &= \sum_{m'=-2}^{+2}[A^{2m'}_{ij}]^{\rm R}D^2_{m'0}(\Omega_{\rm RL}) \\
    &=\sum_{m'=-2}^{+2}[A^{2m'}_{ij}]^{\rm R}d^2_{m'0}(\beta_{\rm ma})\exp\{im'(\omega_r t-\alpha^0_{\rm RL} )\}, \nonumber
\end{align}
where it is important to note that $d^2_{00}(\beta_{\rm ma})=0$ is the special property of the magic angle that removes anisotropic broadening and the dipolar interaction. Subsequently, we have
\begin{align}
&H^{\mathrm{D}}_{\rm rot, MAS}(t)= \nonumber \\& \sum_{m',p=-2}^{+2}\bigg\{
\sum_{i<j}
C^{\mathrm{D}}_{ij}
[A^{2p}_{ij}]^{\rm M}D^{2}_{pm'}(\Omega_{\rm MR})D^2_{m'0}(\Omega_{\rm RL})\mathcal{I}^{20}_{ij}
 \nonumber\\& +
\sum_{m<n}
C^{\mathrm{D}}_{mn}
[A^{2p}_{mn}]^{\rm M}D^{2}_{pm'}(\Omega_{\rm MR})D^2_{m'0}(\Omega_{\rm RL})\mathcal{S}^{20}_{mn}
\label{eq:MAS_secular_dipolar_Hamiltonian}\\& +
\sqrt{\frac{2}{3}}\sum_{i,m}
C^{\mathrm{D}}_{im}
[A^{2p}_{im}]^{\rm M}D^{2}_{pm'}(\Omega_{\rm MR})D^2_{m'0}(\Omega_{\rm RL})\mathcal{I}^{10}_{i}\mathcal{S}^{10}_{m}\bigg\}, \nonumber
\end{align}
as the, now time dependent, dipolar Hamiltonian in the MAS case under the rotating frame and high magnetic field approximation. For a powdered sample, the Euler angles $\Omega_{\rm MR}$ are random variables, and are in practice sampled from the relevant powder averaging scheme.

\begingroup
\renewcommand{\arraystretch}{2}
\begin{table}[htbp]
\centering
\resizebox{\columnwidth}{!}{%
\begin{tabular}{@{}c@{}}
 \hline
 Spherical tensor components \\%[1ex]
 $\mathcal{I}^{\lambda \mu}$ \\
 \hline\hline
 $\mathcal{I}^{10}_i = I_{iz} = \frac{1}{2}Z_i$ \\%[5ex]
 $\mathcal{I}^{1\pm1}_i = \mp\frac{1}{\sqrt{2}}I^{\pm}_i= \mp\frac{1}{2\sqrt{2}}(X_i \pm iY_i)$ \\%[5ex]
 $\mathcal{I}^{20}_{ij} = \frac{1}{\sqrt{6}}(3I_{iz}I_{jz} - \vec{I}_i\cdot\vec{I}_j) = \frac{1}{4\sqrt{6}}(3Z_iZ_j - \vec{\sigma}_i\cdot\vec{\sigma}_j)$ \\%[5ex]
 $\mathcal{I}^{2\pm1}_{ij} = \mp\frac{1}{2}(I^{\pm}_{i}I_{jz} + I_{iz}I^{\pm}_j) = \mp\frac{1}{8}((X_i\pm iY_i)Z_{j} + Z_i(X_j \pm iY_j))$ \\%[5ex]
 $\mathcal{I}^{2\pm2}_{ij} = \frac{1}{2}I^{\pm}_iI^{\pm}_j = \frac{1}{8}((X_iX_j-Y_iY_j) \pm i(X_iY_j + Y_iX_j))$ \\%[1ex]
 \hline
\end{tabular}%
}
\caption{Relevant single spin and two spin spherical tensor components $\mathcal{I}^{\lambda\mu}$ expressed in their cartesian spin operator and Pauli form.}
\label{table:spherical_tensors}
\end{table}
\endgroup

%%%%%%%%%%%%%%%%%%%%%%%%%%%%%%%%%%%%%%%%%%%%%%%%%%%%%%%%%%
\subsection{Hamiltonian Design Techniques}\label{app:Ham_design}
%%%%%%%%%%%%%%%%%%%%%%%%%%%%%%%%%%%%%%%%%%%%%%%%%%%%%%%%%%
We now detail the two ways effective Hamiltonians of the form (\ref{target}) can be engineered, for the case of a static powder and a powder undergoing magic angle spinning.

\subsubsection{Radio-Frequency Pulse Design on Static Samples}

The application of time-dependent control fields provides a means of engineering effective spin Hamiltonians by selectively modifying the evolution generated by the underlying spin interactions. In particular, appropriately designed pulse sequences can suppress, enhance, or rescale individual terms in the Hamiltonian. The resulting dynamics are conveniently described in the interaction, or toggling, frame defined by the control fields~\cite{mehring1983highresolution}
\begin{equation}
\label{eq:Hrot_static_case}
H_{\rm rot}(t) = H^{\mathrm{rf}}_{\rm rot}(t) + H^{\mathrm{int}}_{\rm rot},
\end{equation}
where the internal Hamiltonian is time independent due to the sample being static, we decompose the corresponding propagator as
\begin{equation}
U_{\rm rot}(t,t_a) = U^{\mathrm{rf}}_{\rm rot}(t,t_a)\tilde{U}(t,t_a),
\end{equation}
where $U^{\mathrm{rf}}_{\rm rot}$ describes the evolution due to the applied radio-frequency fields. The remaining propagator $\tilde{U}$ is then determined by the toggling-frame Hamiltonian
\begin{equation}
\tilde{H}(t)
=
U^{\mathrm{rf},\dagger}_{\rm rot}(t,t_a)
H_{\mathrm{int}}
U^{\mathrm{rf}}_{\rm rot}(t,t_a).
\end{equation}
Thus, while the intrinsic spin interactions are unchanged in the original frame, the control sequence periodically modulates their influence in the toggling frame. For a cyclic pulse sequence with cycle time $\tau_c$, the control propagator satisfies the cyclic condition
\begin{equation}
U_{\mathrm{rf}}(t_a+\tau_c,t_a) = \mathbbm{1},
\end{equation}
so that, at the beginning and end of each cycle, the toggling-frame propagator coincides with the full propagator. This allows the dynamics to be described stroboscopically in terms of an effective Hamiltonian. Average Hamiltonian theory provides a systematic construction of this effective Hamiltonian by replacing the time-dependent toggling-frame Hamiltonian with a time-independent operator whose propagator reproduces the exact evolution over one cycle. Thus,
\begin{equation}
\tilde{U}(t_a+\tau_c,t_a)
=
\exp\left(-i\bar{H}\tau_c\right),
\end{equation}
where $\bar{H}$ is the average Hamiltonian associated with the pulse sequence. The average Hamiltonian $\bar{H}$ can be calculated perturbatively using the Magnus/high frequency expansion~\cite{mehring1983highresolution,bukovUniversalHighfrequencyBehavior2015}. The first order Magnus term has the form
\begin{equation}
    \bar{H}^{(1)} = \frac{1}{\tau_c}\int_0^{\tau_c}\tilde{H}(t)dt = \frac{1}{\tau_c}\int_0^{\tau_c} 
    U^{\mathrm{rf},\dagger}_{\rm rot}(t)
    H_{\mathrm{int}}
    U^{\mathrm{rf}}_{\rm rot}(t) dt.
    \label{First_order_Magnus}
\end{equation}
For the purposes of this work, we restrict the effective Hamiltonian to first order in the Magnus expansion. Higher-order Magnus terms provide corrections to this effective description, which are assumed to be negligible.

In NMR, spherical tensor operators provide a convenient representation for Hamiltonian engineering, since radio-frequency fields act primarily by rotating the spin operators. These rotations can be parameterized by the Euler angles $\Omega^{I/S}_{\mathrm{rf}}(t)$, allowing the transformation of the individual spherical-tensor components to be tracked directly. Starting from the secular dipolar Hamiltonian (\ref{eq:secular_dipolar_Hamiltonian}) and applying the first-order average-Hamiltonian construction, the effective interaction can be written as
\begin{align}
\bar{H}^{(1)} &=
\sum_{i<j}
C^{\mathrm{D}}_{ij}
[A^{20}_{ij}]^{\rm L}(\sum_{\mu}\lambda^{II}_{2\mu}\mathcal{I}^{2\mu}_{ij}) \nonumber
\\&+
\sum_{m<n}
C^{\mathrm{D}}_{mn}
[A^{20}_{mn}]^{\rm L}(\sum_{\nu}\lambda^{SS}_{2\nu}\mathcal{S}^{2\nu}_{mn})
\\& +
\sqrt{\frac{2}{3}}\sum_{i,m}
C^{\mathrm{D}}_{im}
[A^{20}_{im}]^{\rm L}(\sum_{\mu,\nu}\lambda^{IS}_{1\mu,1\nu}\mathcal{I}^{1\mu}_{i}\mathcal{S}^{1\nu}_{m}) \nonumber
\end{align}
The coefficients $\lambda$ are determined by the applied pulse sequence and specify the weights with which the different spherical-tensor components contribute to the average Hamiltonian. By choosing the RF rotations such that the secular dipolar terms are selectively transformed to match our target Hamiltonian, Eq.~(\ref{target}), we obtain
\begin{align}
    \bar{H}^{(1)} &= \lambda^{II}\sum_{i<j} C^{\mathrm{D}}_{ij}
    [A^{20}_{ij}]^{\rm L}(\mathcal{I}^{2+2}_{ij}+\mathcal{I}^{2-2}_{ij}) \nonumber \\ 
    &+ \lambda^{SS}\sum_{m<n} C^{\mathrm{D}}_{mn}
    [A^{20}_{mn}]^{\rm L}(\mathcal{S}^{2+2}_{mn}+\mathcal{S}^{2-2}_{mn})\\
    &+ \lambda^{IS}\sqrt{\frac{2}{3}}\sum_{i,m}C^{\mathrm{D}}_{im}
    [A^{20}_{im}]^{\rm L}\mathcal{I}^{10}_{i}\mathcal{S}^{10}_{m} \nonumber
\end{align}
The corresponding nonzero scaling factors are defined by
\begin{align}
    &\lambda^{II} 
    \equiv 
    \Re(\lambda^{II}_{2,+2}) = \Re(\lambda^{II}_{2,-2}), \nonumber\\
    &\lambda^{SS} 
    \equiv 
    \Re(\lambda^{SS}_{2,+2}) = \Re(\lambda^{SS}_{2,-2}),
    \label{eq:define_scaling_factors}\\
    &\lambda^{IS} 
    \equiv 
    \lambda^{IS}_{(1,0),(1,0)}  \nonumber
\end{align}
and are taken to be real. They are determined by the two sets of time-dependent Euler angles associated with the RF fields, $\Omega_{\mathrm{rf}}^{I}(t)$ and $\Omega_{\mathrm{rf}}^{S}(t)$, according to
\begin{align}
     \lambda^{IS}_{(1,0),(1,0)} =& \frac{1}{\tau_c}\int_0^{\tau_c}D^{1}_{00}(\Omega_{\mathrm{rf}}^{I}(t')^{-1})D^{1}_{00}(\Omega_{\mathrm{rf}}^{S}(t')^{-1})dt' \nonumber \\ &= \langle \cos(\beta_{\mathrm{rf}}^{I}(t))\cos{(\beta_{\mathrm{rf}}^{S}(t))}\rangle,
     \label{CH}
\end{align}
\begin{align}
    \lambda^{II}_{2, \pm2} =& \frac{1}{\tau_c}\int_0^{\tau_c}D^{2}_{\pm20}(\Omega_{\mathrm{rf}}^{I}(t')^{-1})dt' \nonumber
    \\&=\sqrt{\frac{3}{8}} \langle \sin^{2}(\beta_{\mathrm{rf}}^{I}(t)) e^{\pm 2i \gamma_{\mathrm{rf}}^{I}(t)} \rangle
    \label{HH},
\end{align}
\begin{align}
     \lambda^{SS}_{2, \pm2} = &\frac{1}{\tau_c}\int_0^{\tau_c}D^{2}_{\pm20}(-\Omega_{\mathrm{rf}}^{S}(t'))dt' \nonumber
     \\&=\sqrt{\frac{3}{8}} \langle \sin^{2}(\beta_{\mathrm{rf}}^{S}(t)) e^{\pm 2i \gamma_{\mathrm{rf}}^{S}(t)} \rangle,
     \label{CC}
\end{align}
where $\langle \phantom{.}\rangle$ denotes a time average over one pulse cycle, and they are labeled by their spin rank and component $(\lambda, \mu)$.

The scaling factors above are defined with respect to the standard normalization of the irreducible spherical tensor operators~\cite{mehring1983highresolution}. To express the resulting Hamiltonian in the Pauli convention used in the main text, we use the explicit relations between the spherical tensor operators and Cartesian spin operators to relate the two sets of scaling factors:
 \begin{align}
  \lambda^{IS} =& 2\eta^{IS} \nonumber\\
    \lambda^{II}  = -\frac{4\eta^{II}}{\sqrt{6}} \quad\quad &
      \lambda^{SS} = -\frac{4\eta^{SS}}{\sqrt{6}}, 
      \label{eq:scaling_relations}
 \end{align}
where $\eta$ denotes the corresponding scaling factors in the Pauli convention.
\subsubsection{MAS Hamiltonian Design}
\label{sec:mas-design}
In the case of MAS, the internal Hamiltonian is time dependent due to the rotation of the sample, Eq.~\eqref{eq:MAS_secular_dipolar_Hamiltonian}, such that the full rotating frame Hamiltonian is
\begin{equation}
    H_{\rm rot}(t) = H^{\rm rf}_{\rm rot}(t) + H^{\rm int}_{\rm rot}(t).
\end{equation}
Over a specified time period, $\tau_c$, we can suspend rotational averaging and generate the operator structure of Eq.~(\ref{target}) via rotor synchronized radio-frequency recoupling sequences. These can selectively reintroduce the desired homonuclear double quantum and heteronuclear longitudinal two spin effective Hamiltonian components, to first order in the Magnus expansion.

Symmetry properties of the interaction frame Hamiltonian, generated by the time-symmetry relations of rf Euler angles with a recoupling sequence, produce selection rules detailing which effective Hamiltonian terms are symmetry allowed. The scaling factors of the symmetry allowed terms depend on the particular recoupling sequence used, and a general symmetry allowed term between two homonuclear spins $(i,j)$ takes the form 
\begin{align}
    \bar{H}^{(1)}_{ij} =\sum_{\mu, m'} C^{\rm{D}}_{ij}\lambda^{II}_{(2,\mu),(2,m')}[A^{2m'}_{ij}]^{\rm R}d^2_{m'0}(\beta_{\rm ma})\nonumber \\
    \times \exp{-im'\alpha^0_{\rm RL}}\mathcal{I}^{2\mu}_{ij}
\end{align}
and the form 
\begin{align}
    \bar{H}^{(1)}_{im} = \sqrt{\frac{2}{3}}\sum_{\mu, \nu, m'}C^{\rm{D}}_{im}\lambda^{IS}_{(1,\mu), (1, \nu),(2,m')}[A^{2m'}_{im}]^{\rm R}d^2_{m'0}(\beta_{\rm ma})\nonumber \\
    \times \exp{-im'\alpha^0_{\rm RL}}\mathcal{I}^{1\mu}_{i}\mathcal{S}^{1\nu}_{m}
\end{align}
for two heteronuclear spins $(i,m)$. The scaling factors, $\lambda$, are labeled by spin rank and component $(2, \mu)$, and spatial rank and component $(2,m')$. To generate the homonuclear double quantum operator structure of Eq.(\ref{target}), the employed pulse sequence scheme must select four spin-space component combinations $(\mu, m') = (2,\pm m'), (-2, \pm m')$ on both the I and S channels. Choosing multiple spatial components for each spin component would generate a so-called ``non-gamma encoded" Hamiltonian~\cite{doi:https://doi.org/10.1002/9780470034590.emrstm0551}, meaning the Hamiltonian amplitudes depend on both angles $(\beta^{\rm R}, \gamma^{\rm R})$. It is also possible to engineer a ``gamma encoded" Hamiltonian, where the amplitudes only depend on $\beta^{\rm R}$, which generates qualitatively similar OTOC dynamics. Along with this choice, we can then also select $m'$ to be 1 or 2 via specific pulse sequence design. 

To maximize OTOC decay speed it is preferable to select a non-gamma encoded homonuclear Hamiltonian, and $m'=1$. 
We also require the heteronuclear spin interactions to have $\mu, \nu=0$, with the same $m'$ as the homonuclear channels. These specific spatial and spin components across two channels must be selected simultaneously, whilst also keeping other unwanted interactions suppressed - requiring a careful choice of recoupling sequence. The resulting scaling factors for the MAS case
are given by
\begin{align}
    &\lambda^{IS}_{(1,0),(1,0),(2,m')} = \langle d^{1}_{00}(-\beta_{\rm rf}^{I}(t))d^{1}_{00}(-\beta_{\rm rf}^{S}(t))e^{im'\omega_r t} \rangle \\
&\lambda^{II}_{(2,\pm2), (2,m')} = \langle d^{2}_{\pm20}(-\beta_{\rm rf}^{I}(t))e^{i(\pm2 \gamma^{I}_{\rm rf}(t)+m'\omega_r t)}\rangle  \\
&\lambda^{SS}_{(2,\pm2), (2,m')} =\langle d^{2}_{\pm20}(-\beta_{\rm rf}^{S}(t))e^{i(\pm2\gamma^{S}_{\rm rf}(t)+m'\omega_r t)}\rangle.
\end{align}
The spatial components Eq.(\ref{eq:spatial_components}) are
\begin{equation}
    [A^{2m'}_{ij}]^{\rm R} = \sqrt{6}d^2_{m'0}(\beta^{{\rm R}}_{ij})e^{im'\gamma^{\rm R}_{ij}}.
\end{equation}
Defining the scaling factors 
\begin{align}
    \lambda^{II/SS}_{(2, \pm2), (2,\pm m')} \equiv \lambda^{II/SS}_{m'} \in \mathbb{R} \\
    \lambda^{IS}_{(1,0), (1,0), (2,\pm m')} \equiv \lambda^{IS}_{m'} \in \mathbb{R}
\end{align}
we have the non-gamma encoded effective Hamiltonian  
\begin{align}
    \bar{H}^{(1)}_{\rm MAS, m'} &= \bigg \{ 2\sqrt{6}\lambda^{II}_{m'}\sum_{i<j}C^{\rm D}_{ij}d^2_{m'0}(\beta_{ij}^{\rm R})d^2_{m'0}(\beta_{\rm ma}) \nonumber \\& \times \cos\{m'(\gamma^{\rm R}_{ij}-\alpha^0_{\rm RL}) \}(\mathcal{I}^{2+2}_{ij} + \mathcal{I}^{2-2}_{ij}) \bigg \}\nonumber \\ + &\bigg \{2\sqrt{6}\lambda^{SS}_{m'}\sum_{m<b}C^{\rm D}_{mb}d^2_{m'0}(\beta_{mb}^{\rm R})d^2_{m'0}(\beta_{\rm ma})\label{eq:non-gamma encoded Heff}\\& \times\cos\{m'(\gamma^{\rm R}_{mb}-\alpha^0_{\rm RL}) \}(\mathcal{S}^{2+2}_{mb} + \mathcal{S}^{2-2}_{mb}) \bigg \}\nonumber \\ \nonumber 
    + & \bigg \{ 4\lambda^{IS}_{m'}\sum_i\sum_m C^{\rm D}_{im}d^2_{m'0}(\beta_{im}^{\rm R})d^2_{m'0}(\beta_{\rm ma})\\ & \times\cos\{m'(\gamma^{\rm R}_{im} - \alpha^0_{\rm RL})\}\mathcal{I}^{10}_i \mathcal{S}^{10}_m\bigg \}, \nonumber
\end{align}
where we can make the relations in Eq.~\eqref{eq:scaling_relations}  to recover the target Hamiltonian Eq.~(\ref{target}), with the coupling constants defined in the main text in Eq.~(\ref{eq:mas_dipolar_couplings}).

%%%%%%%%%%%%%%%%%%%%%%%%%%%%%%%% 
\section{Scaling Factor Bounds} \label{sec:scaling factor bounds}
%%%%%%%%%%%%%%%%%%%%%%%%%%%%%%%% 
 \subsection{General Bound on Scaling Factors for Hamiltonians of the form (\ref{target})}

The following analysis is carried out assuming the Hamiltonian in the static case, as presented in the main text. In the case of a spinning sample, the presented results follow as an upper bound, however saturation is not always attainable.

\begin{figure}
\centering
\includegraphics{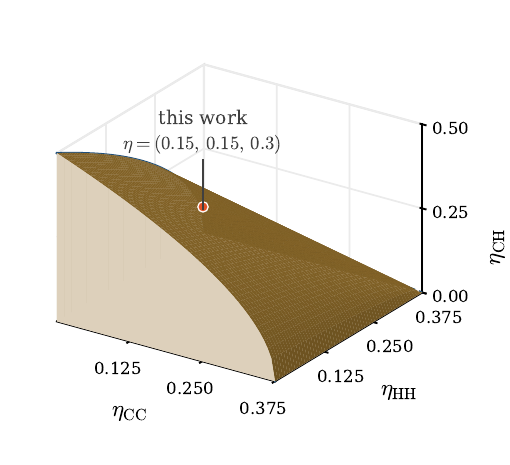}
\caption{Three-dimensional region of physically admissible scaling factors.}
\label{fig:Admissible_Region}
\end{figure}

The scaling factors $\lambda^{IS}$, $\lambda^{II}$, and $\lambda^{SS}$ defined in Eqs.~\eqref{CH}--\eqref{CC} cannot vary independently, since they are determined by physically realizable radio-frequency rotations. Following Ref.~\cite{SALAGER201111}, one may derive bounds on the scaling factors attainable by any RF pulse sequence. To this end, it is useful to consider the time average over one pulse cycle as an inner product on $L^2([0,\tau_c])$,
\begin{equation}
\langle f,g\rangle
\equiv
\frac{1}{\tau_c}
\int_0^{\tau_c} f(t)^*g(t)\,dt.
\end{equation}
The Cauchy--Schwarz inequality then states
\begin{equation}
|\langle f,g\rangle|
\leq
\sqrt{\langle f,f\rangle\langle g,g\rangle}.
\end{equation}
Applying this inequality to $\lambda^{\mathrm{IS}}$ gives
\begin{equation}
|\lambda^{IS}|
=
\left|
\left\langle
\cos\beta_{\rm rf}^{I},
\cos\beta_{\rm rf}^{S}
\right\rangle
\right|
\leq
\sqrt{
\left\langle\cos^2\beta_{\rm rf}^{I}\right\rangle
\left\langle\cos^2\beta_{\rm rf}^{S}\right\rangle
}.
\end{equation}
Using the expressions for $\lambda^{II}$ and $\lambda^{SS}$ in Eqs.~\eqref{HH} and \eqref{CC}, the corresponding averages can be bounded in terms of these scaling factors, yielding
\begin{equation}
|\lambda^{IS}|
\leq
\sqrt{
\left(
1-\sqrt{\frac{8}{3}}|\lambda^{II}|
\right)
\left(
1-\sqrt{\frac{8}{3}}|\lambda^{SS}|
\right)
}.
\end{equation}
Using the relations between the $\lambda$- and $\eta$-conventions, this constraint becomes
\begin{equation}
   2 |\eta^{IS}| \leq \sqrt{( 1-\frac{8}{3}|\eta^{II}|)( 1-\frac{8}{3}|\eta^{SS}|)}.
   \label{bound}
\end{equation}
The homonuclear scaling factors are bounded by $\vert\eta^{II}\vert,\vert\eta^{SS}\vert\leq \tfrac{3}{8}$, so that Eq.~\eqref{bound} defines a three-dimensional region of physically admissible effective (first-order) Hamiltonians of the form of Eq.~\eqref{target}. The resulting admissible region is shown in Fig.~\ref{fig:Admissible_Region}.

\subsection{Symmetric double quantum Hamiltonians}
\label{sec:symm_dq_Hamiltonian}
 
For the simulations in the main text we consider symmetric scaling factors of the form
\begin{equation}
 \eta \equiv \eta_{\mathrm{CC}} = \eta_{\mathrm{HH}} = \frac{\eta_{\mathrm{CH}}}{2},  
 \label{constraint}
\end{equation}
with $0\leq\eta\leq\tfrac{3}{20}$. This choice is motivated by previously engineered Hamiltonians via the pulse sequences TARDIS-I, II \cite{zhang2025quantum}, meaning these are realistic achievable choices. We expect scaling factors that saturate the bound (\ref{bound}) to produce the fastest OTOC decay, leading to the scaling factors given by equation~(\ref{optimal sfs}) in the main text.
To test the optimality of such symmetric scaling factors with respect to OTOC decay time more generally, we considered the protein-RNA system Roquin-mRNA (NA-0), the protein-ligand system Hsp90 (CS-0), and the protein-protein system Fab (AA-06). These were all taken from dataset-C - with a spin network size chosen for each system such that the OTOC is converged. To alleviate simulation time constraints, only one orientation of each system was investigated, which was chosen to best approximate the powder average. Motivated by previous work \cite{zhang2025quantum}, we also consider small systems of terminally $^{13}\mathrm{C}$-labeled alkane chains. We first consider the two-dimensional slice of the admissible region
defined by
\begin{equation}
    \eta_{\mathrm{DQ}}
    \equiv
    \eta_{\mathrm{CC}}
    =
    \eta_{\mathrm{HH}}.
\end{equation}
Within this slice, the admissibility
condition in Eq.~\eqref{bound} reduces to
\begin{equation}
    2|\eta_{\mathrm{CH}}|
    \leq
    1-\frac{8}{3}|\eta_{\mathrm{DQ}}|.
\end{equation}
Defining the OTOC speed as the inverse of the time required for the OTOC to decay to $75\%$ of its initial value, we investigate how the OTOC speed varies across this two-dimensional region of physically admissible scaling factors. Note that in these plots zero speed is assigned to scaling factors that produced OTOCs that did not decay to $75\%$ of its initial value within a realistic experimental window. This does not alter the interpretation of the plots as we are mainly concerned with the optimal fast decay region.

\begin{figure}
    \centering
    \includegraphics[width=\columnwidth]{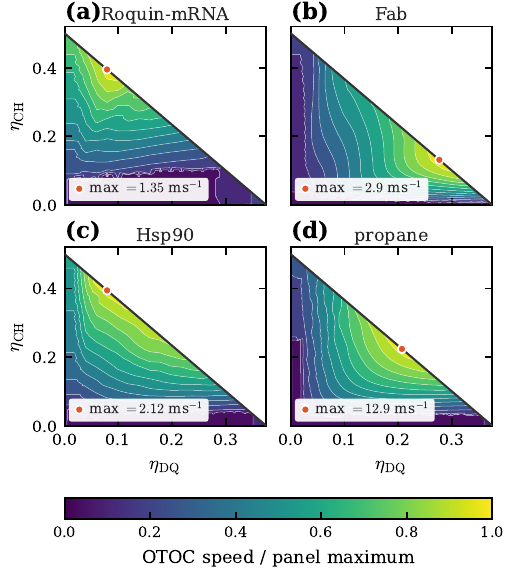}
    \caption{OTOC speed vs. scaling factor for the slice of the admissible region with symmetric double quantum blocks, for four test systems. The optimal scaling factors for OTOC decay are marked. }
    \label{fig:speed_sweep_symmetric}
\end{figure}

\begin{figure}
    \centering
    \includegraphics{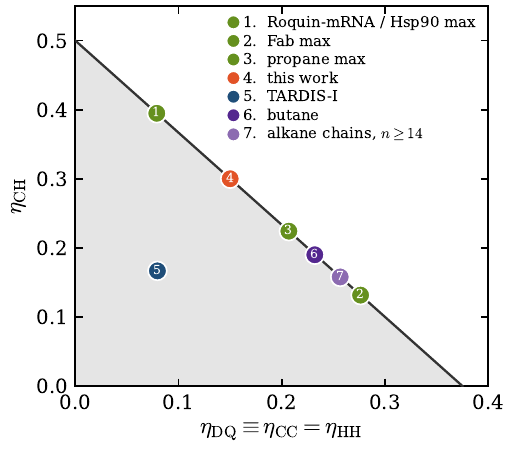}
    \caption{Optimal scaling factors for OTOC speed with the symmetric double quantum constraint for our test proteins and alkane chain systems compared to the scaling factors in TARDIS-I and the ones used in this paper.}
    \label{fig:all_boundary}
\end{figure}

As an illustrative example, Fig.~\ref{fig:speed_sweep_symmetric} shows the OTOC speed across the symmetric slice of the admissible region for our test proteins, and terminally $^{13}\mathrm{C}$-labeled propane. The fastest OTOC decay occurs on the boundary of the admissible region in all systems investigated. Motivated by this observation, we assume that the optimal scaling factors for all the protein systems likewise lie on the boundary of the admissible region. Fig.~\ref{fig:all_boundary} shows that our chosen scaling factors lie on this boundary, and lie between the optimal scaling factors for the proteins investigated. This suggests that further acceleration of the OTOC decay may be possible by engineering Hamiltonians with scaling factors adapted to the specific system.

For the alkane chains, we find that the optimal scaling factors shift towards larger $\eta^{\mathrm{DQ}}$ and smaller $\eta^{IS}$ as the system size increases. This trend can be understood from the increasing importance of rapid operator spreading through the homonuclear proton network relative to the heteronuclear ``bottleneck'' provided by the $^{13}$C spins, which mediate operator propagation into and out of the network associated with the measured (butterfly) operator. For comparison, the scaling factors realized by TARDIS lie within the admissible region and therefore do not saturate the bound, suggesting further scope for optimization.

\subsection{General double quantum Hamiltonians}\label{app:general_eff_DQHam}

\begin{figure} 
\centering 
\includegraphics[width=\columnwidth]{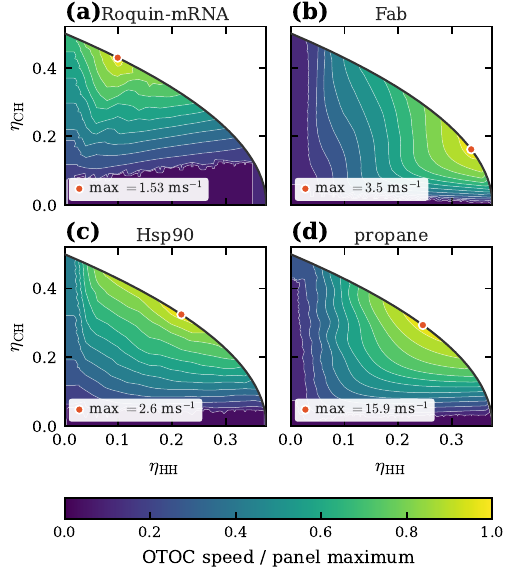} 
\caption{OTOC speed vs. scaling factor for the slice of the admissible region with $\eta_{\mathrm{CC}}=0$ for four test systems. The optimal scaling factors for OTOC decay are marked.} 
\label{fig:Propane_speed_sweep_CC0} 
\end{figure}

For completeness, we also consider the optimal scaling factors without imposing the requirement of symmetric double-quantum interactions, such that $\eta^{II}\neq\eta^{SS}$. Effective Hamiltonians of this more general form could in principle be realized through novel pulse-sequence design or numerical approaches such as optimal control~\cite{KHANEJA2005296}. For the larger systems of interest here, we expect the carbon--carbon double-quantum interaction to make a negligible contribution to the OTOC decay. This was demonstrated in our previous work~\cite{zhang2025quantum}, and is also true for the protein system considered here. This reflects the distinct roles of
the interactions in operator spreading: operator spreading is primarily mediated by the $I$ spin network and its coupling to the $S$ spins, rather than by $S$--$S$ dipolar interactions. We therefore set
$\eta_{\mathrm{CC}}=0$. A sweep over the resulting two-dimensional admissible region, 
\begin{equation} 2|\eta_{\mathrm{CH}}| \leq \sqrt{1-\frac{8}{3}|\eta_{\mathrm{HH}}|}, 
\end{equation} 
again shows that the optimal OTOC speed occurs on the boundary (Fig.~\ref{fig:Propane_speed_sweep_CC0}). Moreover, relaxing the symmetric double-quantum constraint allows for faster OTOC decay than obtained within the symmetric subspace considered above. 
Furthermore, allowing $\eta_{\mathrm{CC}}$ to increase from zero and reoptimizing the remaining scaling factors leads in general to a reduction in the OTOC speed for the systems considered here. This behavior is consistent with the admissibility constraint: reducing $\eta^{SS}$ increases the available range of $\eta^{II}$ and $\eta^{IS}$, allowing greater weight to be placed on the interactions that most directly contribute to operator spreading.

\section{Description of protein systems}\label{sec:App_systems}

\subsection{Protein-ligand systems}
\subsubsection{Hsp90}
The structure of the Hsp90:CCT018159 protein-ligand complex, which we have abbreviated to ``Hsp90'' throughout, was taken from Ref.~\cite{cheung2005hsp} (PDB code: 2BRC, X-ray diffraction, 1.6~\AA{} resolution). ProtClean~\cite{protclean} was used to standardize the file, add hydrogens and ACE/NME caps, and add missing atoms to 10 residues specified in the original PDB. This cleaned PDB file was used for analysis and running MD trajectories.

Heat Shock Protein 90 (Hsp90) plays a critical role in using ATP to chaperone the folding and stabilizing ``client'' proteins that are involved in cancer development and progression, making it a potential cancer drug target. The ligand 3,4-diarylpyrazole (CCT018159) bound to Hsp90 in this system was found to act as a potent inhibitor of Hsp90 ATPase activity (7.1 $\mu$M).

The central and distal carbons in the ligand were chosen so that some carbon-carbon pairs between nearby protein residues involved in binding interactions and hydrogen bonding with the ligand would be included in the range between 5-15~\AA{}. These include Ile82, Met84, Leu93, Phe124, and Leu173 for binding residues and Gly97 for hydrogen bonding residues.

\subsubsection{A2AR-ZMA}
The structure of the A$_{2A}$AR-BRIL-$\Delta$C:ZM241385 protein-ligand complex, which we have abbreviated to ``A2AR-ZMA'' throughout, was taken from Ref.~\cite{liu2012a2ar} (PDB code: 4EIY, X-ray diffraction, 1.80~\AA{} resolution). ProtClean was used to standardize the file and add hydrogens and ACE/NME caps to the protein's termini.  ProtClean also added 15 residues missing in the initial PDB (Ala1043:Glu1057). An additional 42 located at the termini were not added.  Missing atoms were also added to Met1058 and Lys1059. This cleaned PDB file was used for analysis and running MD trajectories.

The human A$_{2A}$ adenosine receptor (A$_{2A}$AR) is a G protein-coupled receptor (GPCR). GPCRs transmit signals through cell membranes, making them useful as drug targets. The crystal structure provided in Ref.~\cite{liu2012a2ar} is of a mutated version of the wild type protein, with its third intracellular loop replaced with apocytochrome b$_{562}$RIL (BRIL). BRIL is a mutated helix bundle derived from \textit{Escherichia coli} that helped to stabilize the protein, allowing for a higher resolution crystal structure. This mutated version is referred to as A$_{2A}$AR-BRIL-$\Delta$C.

The central and distal ligand carbons were chosen to be on the central double ring and the five-membered ring deeper in the binding pocket, respectively. The internal water channel formed by the protein has two narrow bottlenecks formed by residues Trp246 and Tyr288. Trp246 is included in the carbon-carbon pairings for all combinations.

\subsubsection{TYK2-cmpd8}
The structure of the TYK2:compound-8 protein-ligand complex, which we have abbreviated to ``TYK2-cmpd8'' throughout, was taken from Ref.~\cite{fensome2018tyk2} (PDB code: 6DBK, X-ray diffraction, 2.00~\AA{} resolution). ProtClean was used to standardize the file and add hydrogens and ACE/NME caps to the protein's termini. There were also 6 missing atoms in Phe908 that ProtClean remedied. This cleaned PDB file was used for analysis and running MD trajectories.

Tyrosine Kinase 2 (TYK2) is an enzyme in the Janus kinase (JAK) family that sends immune and inflammatory signals to proteins such as Interleukin-23 (IL-23). TYK2 works with other JAK members like JAK1. Ref.~\cite{fensome2018tyk2} synthesized a dual TYK2/JAK1 inhibitor to provide increased efficacy for various inflammatory diseases, such as psoriasis and arthritis. Compound-8 is one such selective inhibitor that prevents excessive pro-inflammatory signaling by binding to the protein's ATP site and is the precursor to the highly effective compound 23 (also known as brepocitinib).

The ligand carbons were chosen to be on the diaminopyrimidine ring (central) that is highly involved with docking with TYK2 and near the constrained bridge (distal) that helps to reduce the molecule's flexibility.

\subsection{Protein-protein systems}
\subsubsection{Fab}
The structure of the KcsA:Fab protein-protein complex, referred to as ``Fab'' throughout because the interface of interest is between the heavy and light chains of the Fab antibody fragment, was taken from Ref.~\cite{zhou2001kcsa} (PDB code: 1K4C, X-ray diffraction, 2.0~\AA{} resolution). ProtClean was used to standardize the file while adding hydrogens and ACE/NME caps to the proteins' termini. The original PDB included a tetramer of KcsA:Fab units. One monomer consists of heavy and light chains of Fab along with a complexed KcsA protein. A single KcsA:Fab monomer using the original PDB's chains A and B for Fab and chain C for KcsA was used for the MD trajectories in this paper because the formation of the ion channel was not important to the OTOC analysis of the interaction between the heavy and light chains of Fab. The original PDB also included diacyl glycerol, nonan-1-ol, and K$^{+}$ ions. The alcohol and ions were included in the cleaned structure while the glycerol was left out.

The KcsA potassium channel is commonly complexed with monoclonal Fab antibody fragments to act as a chaperone, stabilizing the channel and promoting crystal lattice contacts to allow X-ray crystallography. Fab consists of two distinct chains, a heavy and light chain, which bind to each other to create a wide interface region. Both chains also simultaneously bind to one end of the KcsA protein. The Fab fragment is the antibody-binding region of the larger antibody and binds to KcsA in such a way to stabilize the KcsA tetramer while still allowing potassium to move freely through the channel that is formed.

The carbon-carbon pairings were chosen to be located at the interface of the Fab antibody's heavy and light chains as their interface is large and less localized than the interface between KcsA and Fab. This allowed for a wider distribution of distances across a larger region to be analyzed.

\subsubsection{KRas-R11}
The structure of the KRasWT:R11.1.6 protein-protein complex, referred to as ``KRas-R11'' throughout, was taken from Ref.~\cite{kauke2017kras} (PDB code: 5UFE, X-ray diffraction, 2.30~\AA{} resolution). ProtClean was used to standardize the file and add hydrogens and ACE/NME caps to the proteins' termini. The initial PDB had 7 missing residues in R11.1.6, of which 4 residues were added by ProtClean. The 3 remaining residues were at the protein's C-terminus and were not included. Twelve additional residues had missing atoms, all of which were added by ProtClean. The GNP (non-hydrolyzable GTP analog) ligand and the Mg$^{2+}$ ion were retained in the cleaned structure.

KRas is an important GTPase playing key roles in cell growth and proliferation in related pathways. Its mutations, however, can cause uncontrolled tumor growth. The R11.1.6 protein binds to both wild-type KRas and some of its mutants, providing structural information on how the protein conformations affect specificity.

\subsubsection{TRalpha-RXR}
The structure of the TR$\alpha$:RXR protein-protein complex, referred to as ``TRalpha-RXR'' throughout, was taken from Ref.~\cite{putcha2012trrxr} (PDB code: 3UVV, X-ray diffraction, 2.95~\AA{} resolution). ProtClean was used to standardize the file and add hydrogens and ACE/NME caps to the proteins' termini. The initial PDB had 15 missing residues in thyroid receptor $\alpha$ (TR$\alpha$), of which 9 were added by ProtClean (the others were located at the termini and not added). There were also 26 missing residues in retinoid X receptor (RXR), 14 of which were added by ProtClean. TR$\alpha$ also had missing atoms in 9 residues, all of which were added. The crystal structure also included the bound ligand 9-\textit{cis} retinoic acid, which was retained in the cleaned structure and subsequent MD trajectories, and the thyroid hormone 3,3',5 triiodo-\textsc{l}-thyronine (T3), which was not included.

Thyroid hormones control development and metabolism in mammals, which are regulated by thyroid hormone receptors, or TRs. Some receptors function as homodimers, while others, including the TR:RXR complex, are present as heterodimers. The introduction of 9-\textit{cis} retinoic acid (9c) to the complex induces allosteric communication between 9c and TR$\bullet$T3, resulting in increased local dynamics in the TR binding pocket. The protein-ligand interactions of each protein with its respective ligand likely warrant additional study.

\subsection{Protein-DNA/RNA systems}
\subsubsection{ETV6-DNA}
The structure of the ETV6:DNA protein-DNA complex, referred to as ``ETV6-DNA'' throughout, was taken from Ref.~\cite{de2014etv6} (PDB code: 4MHG, X-ray diffraction, 2.2~\AA{} resolution). ProtClean was used to standardize the file while adding hydrogens and ACE/NME caps to the protein's termini. This cleaned PDB file was used for analysis and running MD trajectories.

This system consists of the ETS transcriptional repressor ETV6 bound to a 14 base-pair DNA duplex. The specific DNA strand contains a GGAA sequence that is recognized by the protein, leading to high affinity. An alpha-helix known as Helix 5 blocks the binding site in free inhibited ETV6. This helix is not included in the crystal structure due to its disordered nature. Not counting the missing helix, the structure of the bound ETV6 used in this work is close to a previously determined free inhibited ETV6.

The base and backbone DNA carbons were chosen to be near the remaining helices in the protein. This was both to ensure as broad a range of pairs with protein carbons could be found, as well as to make pairs with important contacts in the protein, such as residues Leu337 and Trp376 with the DNA backbone and residues His396 and the sidechain of Arg392 with DNA bases.

\subsubsection{TP901-1-DNA}
The structure of the TP901-1:DNA protein-DNA complex, referred to as ``TP901-1-DNA'' throughout, was taken from Ref.~\cite{frandsen2013tp901} (PDB code: 3ZHM, X-ray diffraction, 2.60~\AA{} resolution). ProtClean was used to standardize the file and add hydrogens and ACE/NME caps to the protein's termini. This cleaned PDB file was used for analysis and running MD trajectories.

Bacteriophages like TP901-1 are a significant problem in dairy production, as unsterilized milk can contain these phages, potentially resulting in entire starter cultures being wiped out. This crystal structure consists of the helix-turn-helix motif containing N-terminal domain of the CI repressor protein of TP901-1 complexed with a 9 base-pair DNA sequence.

While TP901-1 binding to DNA results in no major conformational changes, some residues have small movements, such as Gln51, which then interacts with Gln45. Both of these residues were included (alpha and side chains) in the carbon-carbon pairs with backbone and base DNA carbons.

\subsubsection{PABP-RNA}
The structure of the PABP:RNA protein-RNA complex, referred to as ``PABP-RNA'' throughout, was taken from Ref.~\cite{deo1999pabp} (PDB code: 1CVJ, X-ray diffraction, 2.60~\AA{} resolution). ProtClean was used to standardize the file, and add hydrogens and ACE/NME caps to the protein's termini. The crystal structure in the initial PDB file consisted of 8 subunits of bound PABP-RNA complexes, of which one was selected for analysis and molecular dynamics simulations.  The adenosine monophosphate (AMP) was not included in the molecular dynamics simulations.

Human poly(A)-binding protein (PABP) first identifies pre-messenger RNA through sequence recognition followed by binding a poly(A) tail (consisting of up to 250 nucleotides) to the 3' end. This structure plays a critical role in initiating translation in eukaryotes as well as stabilizing mRNA against premature degradation.

PABPs contain RNA-binding domains known as RNA recognition motifs (RRMs) due to their prevalence throughout more than 200 bacterial, fungal, plant, and animal proteins. The PABP sequence used contains two of these RRMs, RRM1 and RRM2, binding the 3' and 5' ends of the mRNA, respectively. Carbons on the base and backbone were chosen near the 5' end in Ade-2 and Ade-3, resulting in pairing with carbons entirely in the RRM2 region.

\subsubsection{Roquin-mRNA}
The structure of the Roquin:mRNA protein-mRNA complex, referred to as ``Roquin-mRNA'' throughout, was taken from Ref.~\cite{tan2014roq} (PDB code: 4QIL, X-ray diffraction, 2.9~\AA{} resolution). The initial PDB contained a dimer of the crystallized bound system. ProtClean was used to select one of the monomers, standardize the file, and add hydrogens and ACE/NME caps to the protein's termini. The initial PDB had 14 missing residues in the protein (Pro112:Thr125) belonging to a disordered loop and a missing OG atom in Ser88, which ProtClean remedied. Other missing residues at the protein's termini (Met86:Met87 and Lys401:His415) were not added. This cleaned PDB file was used for analysis and running MD trajectories.

The Roquin protein (and Roquin-2) plays a key role in repressing autoimmune diseases, with mice deficient in the proteins displaying lupus-like symptoms. Here, the ROQ domain of human Roquin is in complex with the mouse \textit{Hmgxb3} 19-nucleotide stem-loop RNA, referred to as Hmg\textsubscript{19}.

Due to the structure of Roquin, which contains three domains, one of which is situated far from Hmg\textsubscript{19}, the carbon-carbon pairs are also located only in Domains II and III. The base and backbone RNA carbons were selected in order to provide a range of pairs with protein carbons when possible, while also attempting to make pairs with important contacts in the protein, such as residues Lys220 and Arg219 with the RNA backbone and residues Gln247 and the Tyr250 with RNA bases.

\section{Local environment proxy for structural parameter investigation}
\label{sec:toy_model}
The analysis in Sec.~\ref{sec:results_information} uses a deliberately small, fully controlled spin system as a proxy for the local environment of a labeled carbon pair in a protein-ligand system. The aim is not to exactly reproduce any particular protein structure, but to make the OTOC's sensitivity to parameter perturbations computationally tractable.

\subsection{The proxy system}

\begin{figure}
    \centering
    \includegraphics[width=\columnwidth]{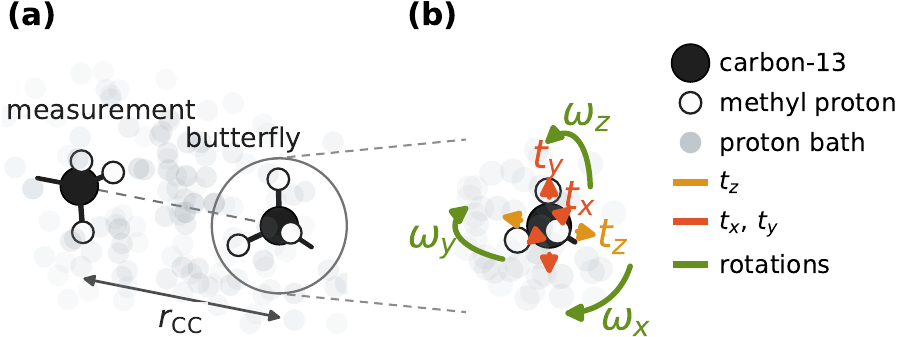}
    \caption{Proxy system used to investigate structural parameters. (a) Schematic illustration of the measurement and butterfly configuration. (b) Geometric perturbations applied to the butterfly unit, assumed to be part of a molecular ligand.} \label{fig:proxy_system_figure}
\end{figure}

The proxy system depicted in Fig.~\ref{fig:proxy_system_figure} consists of two methyl groups (denoted measurement and butterfly), each consisting of a labeled $^{13}\rm{C}$ nucleus and three protons in tetrahedral geometry with $d_{\rm{CH}}=1.09$~\AA, embedded in a further proton bath. These two labeled carbon nuclei serve as measurement and butterfly operators, such that they play the same role as the carbon-carbon pairs in the protein dataset. The separation between the two labeled carbons, $r_{\rm{CC}}$, is varied over $5-15$~\AA, spanning the approximate range of dataset-C.

Throughout, the measurement methyl is held fixed, and the butterfly methyl  - taken to form part of a ligand - is subject to rigid  rotations and translations relative to the other methyl cluster (Fig.~\ref{fig:proxy_system_figure}). The two methyl clusters were given fixed random relative orientations as to remove any point symmetry that could make particular perturbations degenerate, and the internal geometry of both clusters is held rigid throughout.

The total proxy contains $n=12$ spins, two labeled-$^{13}\rm{C}$, six methyl $^1\rm{H}$, and a bath of size $N_{\rm{H}}=4$. The Hamiltonian is Eq.~\eqref{target}, with scaling factors chosen to satisfy the fastest possible OTOC decay choice for symmetric double quantum blocks: $\eta_{\rm{CC}}=\eta_{\rm{HH}}=0.15$ and $\eta_{\rm {CH}}=0.3$ (see Appendix~\ref{sec:symm_dq_Hamiltonian} for details).

For each instance, a large proton bath was generated at random by rejection sampling uniformly within a capsule of radius 9~\AA~around the measurement-butterfly internuclear vector, enclosing the two methyl clusters, at the characteristic proton number density of the proteins investigated in the main text, $\rho_{\mathrm{H}} = 0.06$ \AA$^{-3}$, subject to realistic minimum separation constraints of $d_{\mathrm{HH}} \geq 2$~\AA. Once the protons were placed, they were assigned to each cluster, via coupling strength, in equal numbers, and from this bath the subset of $N_{\rm{H}}=4$ protons coupled the strongest were retained. This selection is made only once, before any perturbations are applied. We do not expect the OTOCs produced by this number of spins to be converged, just to produce similar behaviour under molecular perturbations, over similar timescales, to realistic systems.

As the bath sampling is random, for each $r_{\rm{CC}}$, the analysis is repeated for ten independent bath realizations. The results shown in Sec.~\ref{sec:results_information} are the statistics over these realizations. 

\subsection{Structural parameters and sensitivity analysis}

The rigid rotations and translations of the butterfly methyl cluster are performed around three internal (molecular) axes. We take the molecular z axis to be the internuclear vector connecting the measurement ($^{13}\rm{C}_A$) and butterfly ($^{13}\rm{C}_B$) carbons, and define two more orthogonal molecular axes as
\begin{align}
    \hat{\vec{z}}_{\mathrm {mol}} = \frac{\vec{r}_{^{13}\rm C_{\rm B}} - \vec{r}_{^{13}\rm C_{\rm A}}}{|\vec{r}_{^{13}\rm C_{\rm A}} - \vec{r}_{^{13}\rm C_{\rm B}}|}, \quad & \quad \hat{\vec{x}}_{\rm mol} = \frac{\hat{\vec{x}}_{\rm lab}-(\hat{\vec{x}}_{\rm lab}\cdot\hat{\vec{z}}_{\rm mol})\hat{\vec{z}}_{\rm mol}}{|\hat{\vec{x}}_{\rm lab}-(\hat{\vec{x}}_{\rm lab}\cdot\hat{\vec{z}}_{\rm mol})\hat{\vec{z}}_{\rm mol}|}, \nonumber \\ \hat{\vec{y}}_{\rm mol}=&\hat{\vec{z}}_{\rm mol} \times \hat{\vec{x}}_{\rm mol}. 
\end{align}
The ligand proxy is displaced along, and rotated around these three axes in turn. The corresponding structural parameters are then  collected in
\begin{equation}
   \theta=(t_z, t_x, t_y, \omega_z, \omega_x, \omega_y),
\end{equation}
which defines the parameter space $\Theta \subset\mathbb{R}^6$. We wish to quantify the sensitivity of OTOCs to these structural parameters, which in turn indicates the identifiability of such parameters when trying to solve the inverse problem discussed in Sec~\ref{sec:model_fitting_structure}. Whilst $\Theta$ is a submanifold of the entire parameter space $\{\vec{R}\}$, in our proxy system every other parameter is known exactly - which is in general not the case. Therefore the qualitative conclusions presented in Sec.~\ref{sec:results_information} are a best-case for the full inverse problem.

To quantify sensitivity, we consider the eigenspectrum of the pullback metric on the structural parameter space, which reveals the number and relative importance of parameter combinations to which the OTOC is locally sensitive.

Defining a forward map F, from our parameter space into the observation space of all possible OTOC trajectories, $\mathcal{O} \subset \mathrm{L}^2[0, \mathrm{T}]$, $\mathrm F : \Theta\rightarrow\mathcal{O}$, we define the image $\mathcal{M}=\mathrm F(\Theta) \subset \mathcal{O}$, to be our model manifold. The pullback metric $g=\mathrm{F}^*g_\mathcal{O}$ onto the parameter manifold is thus 
\begin{equation}
    g_{\mu \nu} = \left \langle \frac{\partial C(\cdot,\theta)}{\partial \theta_\mu},\frac{\partial C(\cdot,\theta)}{\partial \theta_\nu} \right \rangle_{\mathrm{L}^2} = \int \frac{\partial C(t,\theta)}{\partial \theta_\mu}\frac{\partial C(t,\theta)}{\partial \theta_\nu} dt,
\label{metric}
\end{equation}
where $C(t, \theta)$ denotes the powder averaged OTOC. 
In practice, we are interested only in the relative sensitivity along different directions in parameter space. We therefore compute the eigenspectrum of
\begin{equation}
    (J^T J)_{\mu \nu} = \sum_i \frac{\partial C(t_i,\theta)}{\partial \theta_\mu}\frac{\partial C(t_i,\theta)}{\partial \theta_\nu}, \quad J_{i\mu} = \frac{\partial C(t_i,\theta)}{\partial \theta_\mu}.
\label{metric2}
\end{equation}
The Jacobian terms are constructed via finite differences by physically perturbing the molecular coordinates in the way described above, such that 
\begin{equation}
    \frac{\partial C(t, \theta)}{\partial\theta_\mu} \approx \frac{C(t, \theta_\mu + \epsilon_\mu)- C(t, \theta_\mu - \epsilon_\mu)}{2\epsilon_\mu}.
\end{equation}
For a uniform time grid, $J^T J$ is proportional to the discretized form of Eq.~(\ref{metric}). Its eigenvalues quantify the sensitivity of the OTOC trajectory along orthogonal directions in structural parameter space: large eigenvalues correspond to directions that produce large changes in the OTOC, whereas small eigenvalues identify comparatively insensitive directions. The corresponding eigenvectors define the structural eigenmodes, i.e., the particular linear combinations of translations and rotations to which the OTOC is sensitive.

In order to compare rotational and translational eigenvalues and vectors, the Jacobian is rescaled such that we are comparing the root mean squared displacement induced by each of the six perturbations. The displacement induced in spin $i$ by a rotation of a small angle $\delta $ is $\delta |\hat{\vec{n}} \times(\vec{r_i}-\vec{r}_{\rm C_{\rm B}})|$. Thus, for a rotation of cluster B around an axis $\hat{\vec{n}}$, the rotational modes must be scaled by 
\begin{equation}
    l_\omega(\hat{\vec{n}})=\bigg [\frac{1}{4}\sum_{i \in \rm{B}}|\hat{\vec{n}} \times(\vec{r}_i-\vec{r}_{\rm C_{\rm B}})|^2\bigg]^{\frac{1}{2}}
\end{equation}
such that the eigenvalues are comparable. This amounts to making the relation
\begin{equation}
    J \rightarrow J\cdot\text{Diag}(1, 1, 1, l_\omega(\hat{\vec{z}}_{\rm mol})^{-1}, l_\omega(\hat{\vec{x}}_{\rm mol})^{-1}, l_\omega(\hat{\vec{y}}_{\rm mol})^{-1})
    \label{eq:rescaling}
\end{equation}
before the pullback metric (\ref{metric2}) is constructed.

To see how parameter sensitivity relates to identifiability, note that the pullback metric \eqref{metric} is closely related to the classical Fisher information matrix (FIM)~\cite{raoInformationAccuracyAttainable1992,transtrumGeometryNonlinearLeast2011}. The FIM quantifies how much information a measurement carries about the parameters of the model. For a single measured OTOC trajectory, the FIM is given by 
\begin{equation}
    F_{\mu\nu} = -\langle \partial_{\theta_\mu}\partial_{\theta_\nu}\log P(\rm{OTOC|\theta)\rangle},
    \label{eq:FIM}
\end{equation}
where $P(\mathrm{OTOC}|\theta)$ is the probability of measuring the data, given our model parameters, $\theta$, and $\langle \cdot \rangle$ is an average over experimental noise. For an OTOC measured at discrete times $\{t_i\}$, of interval $\Delta t$, under additive Gaussian noise of constant standard deviation $\sigma$, such that $\mathrm{OTOC}_i=C(t_i, \theta^*) + \eta_i$, with independent $\eta_i \sim \mathcal{N}(0, \sigma^2)$ and $\theta^*$ the true parameters, the FIM, Eq.~\eqref{eq:FIM}, reduces to 
\begin{equation}
F = \frac{1}{\sigma^{2}}\,J^{T}J \approx \frac{1}{\sigma^2 \Delta t} g,
\end{equation}
where $J$ is the Jacobian of Eq.~\eqref{metric2}, $g$ is the pullback metric Eq.~\eqref{metric}, and the approximation is the replacement of the integral with a sum. The FIM and pullback metric are proportional, hence they have identical eigenvectors, and eigenvalues that are proportional to one another - leading to identical model manifold effective dimensionality Eq.~\eqref{eq:effective_dimension}. For several independent OTOC datasets, containing multiple measurement-butterfly pairs, the total FIM is the sum of the individual ones.

When a single eigenvalue dominates the eigenspectrum, the effective dimension Eq.(\ref{eq:effective_dimension}) of the model manifold approaches unity. Spectra of this kind, with a small number of stiff directions and a low effective dimension, are the defining feature of sloppy models~\cite{transtrumPerspectiveSloppinessEmergent2015}.

\subsection{Numerical details and convergence}

\begingroup
\renewcommand{\arraystretch}{1.25}
\begin{table}[t]
\caption{Numerical parameters used for the proxy calculations.}
\label{tab:proxyparams}
\footnotesize
\begin{tabular}{@{}ll@{}}
\hline\hline
spins        & $n=12$: 2 $^{13}$C, 6 methyl $^{1}$H, 4 bath $^{1}$H \\
geometry     & $d_{\rm CH}=1.09$~\AA; $r_{\rm CC}=5$--$15$~\AA \\
scaling factors  & $\eta_{\rm CC}=\eta_{\rm HH}=0.15$;
               $\eta_{\rm CH}=0.30$ \\
trace        & exact enumeration \\
powder grid  & ${\rm ZCW}_{1154}$ \\
time grid    & $N_t=21$, uniform \\
window       & $t_{\max}=5\,t_{1/2}$  \\
perturbation & $\epsilon_t=0.05$~\AA; $\epsilon_\omega=0.02$~rad \\
realizations & 10 bath realizations per distance \\
\hline\hline
\end{tabular}
\end{table}
\endgroup

Table~\ref{tab:proxyparams} collects the numerical parameters used in this proxy model study. The qualitative results in Fig.~\ref{fig:FIM} were found to be robust to alterations in time window (where the window remained experimentally accessible), and number of points in the uniform time grid, $N_t$.

\begin{figure}
    \centering
    \includegraphics[width=\columnwidth]{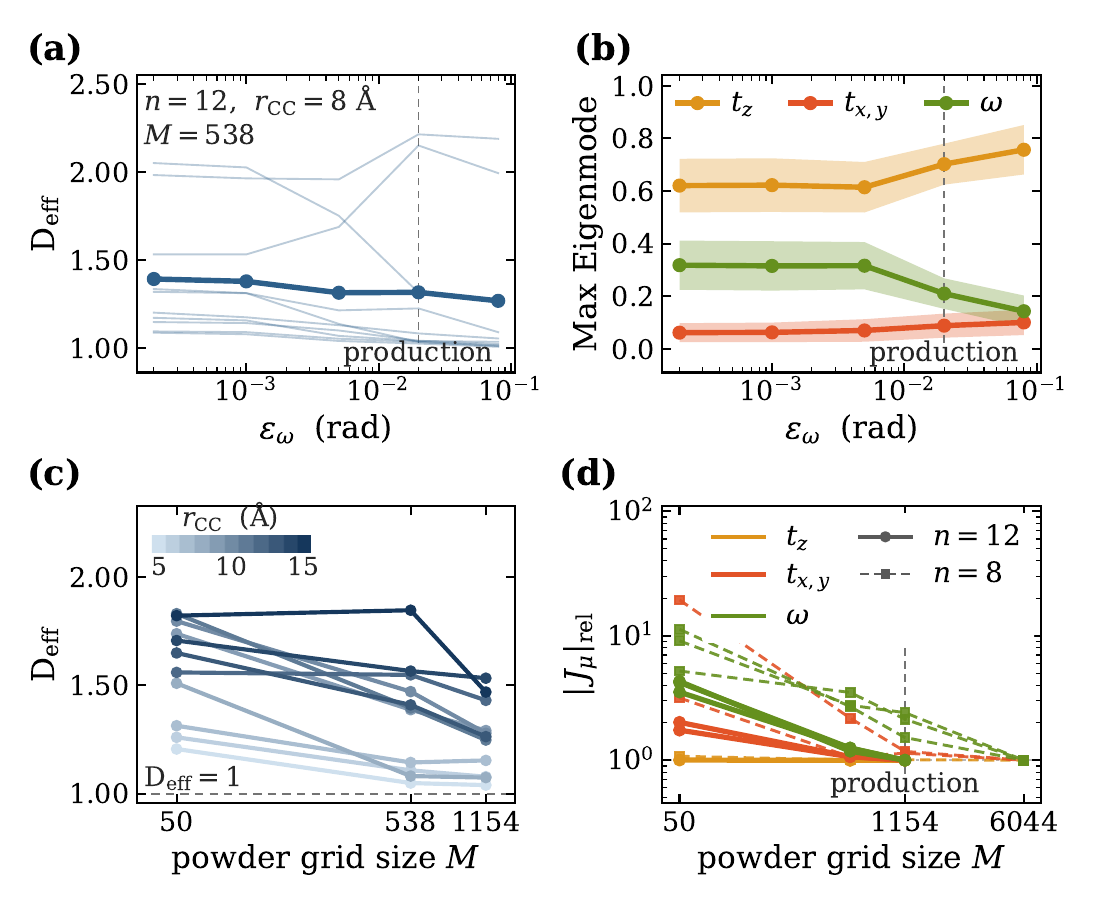}
    \caption{(a) Mean and individual $\mathrm{D}_{\rm eff}$ and (b) leading eigenvector weights for different rotational perturbation size $\epsilon_\omega$ for our proxy model over 10 random proton bath instances, investigated for a smaller ${\rm ZCW}_{538}$ grid. (c) $\mathrm{D}_{\rm eff}$ versus powder grid size, M, for our proxy $n=12$ spin model, this is in general fairly robust but can be seen to be trending towards 1 with M. (d) Variation of the relative Jacobian column magnitudes with powder grid size. These are defined relative to the finest computationally tractable grid, for our proxy model with and without a proton bath such that $|J_{\mu}|_{\rm rel} = |J_{\mu}|/ |J_{\mu}|_{\rm Max\{M\}}$. The Jacobian columns are split into $t_z$, $\{t_x, t_y\}$ and $\omega$, as they are in the main text, demonstrating which modes are robust to the coarsening of the powder grid.}
    \label{fig:proxy_convergence}
\end{figure}

The choices of $\epsilon_t$ and $\epsilon_{\omega}$ are both a numerical and modeling decision, and the effect of varying $\epsilon_\omega$ on one $r_{\rm CC}$ instance is considered in Figs.~\ref{fig:proxy_convergence} (a) and (b). The local proxy model tries to ascertain whether an OTOC can distinguish structures that differ by a given displacement. Thus, the relevant scale of $\epsilon$ is set not by $\epsilon \rightarrow 0$, but by the structural resolution of the given model. The quoted displacements amount to root mean squared displacements of the order $\approx10^{-2}$~\AA. This is one order of magnitude below the coordinate precision of the protein structures of dataset-A, therefore these perturbations probe the smallest possible structural scale we can resolve in this model.

The size of the powder grid has the largest effect on the convergence of these results. Increasing grid size moves the results monotonically towards those reported in Fig.~\ref{fig:FIM}, and we expect larger, computationally expensive grids, to follow this trend and see $\rm{D}_{\rm eff}\rightarrow1$, with an even larger contribution from $t_z$ to the dominant eigenmode. This is investigated for both the proxy model, and the proxy model with the proton bath removed to make larger grids accessible, in Figs.~\ref{fig:proxy_convergence} (c) and (d). The effective dimension and the Jacobian column corresponding to $t_z$ are fairly robust to grid size, however the Jacobian columns corresponding to rotational and transverse translational modes see a large variation. 

\section{Numerical simulation details}

\subsection{Exact simulation error analysis}\label{sec:App_exact_sim_error}
\begin{figure}
    \centering
    \includegraphics[scale=0.7]{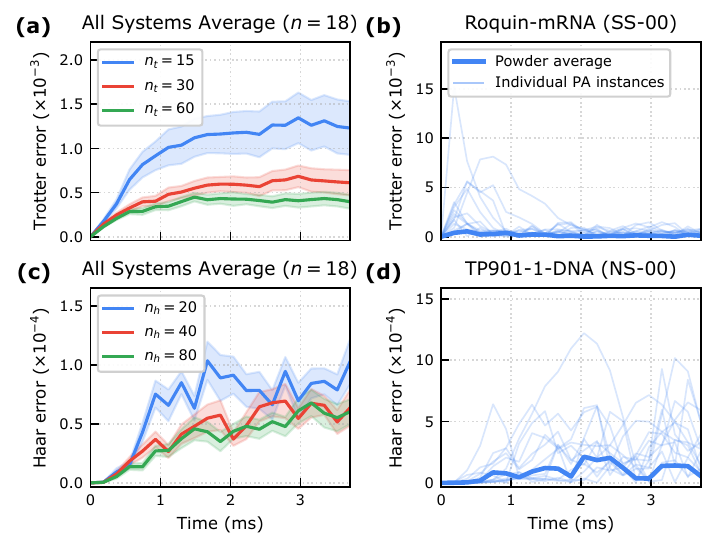}
    \caption{(a) The average Trotter error for $n_t = 15, 30, 60$, computed against an $n_t = 120$ calculation. (b) The Trotter error in the powder averaged OTOC plotted along with the Trotter error for each instance for an example (Roquin-mRNA, SS, 00). (c) The average sampling error for $n_h = 20, 40, 80$ computed against an $n_h = 160$ calculation. (d) The sampling error in the average compared to the sampling error in individual instances for an example (TP901-1-DNA, NS-00). All calculations are done for $n = 18$ spins.}
    \label{fig:error}
\end{figure}

In Fig.~\ref{fig:error}(a) we show the average Trotter error for different number of Trotter steps for 20 carbon-carbon pairs across two protein systems (TYK2-cmpd8 and Roquin-mRNA). The Trotter error is computed vs a reference that uses 120 time steps per time interval. At $n_t = 15$ Trotter steps, the Trotter error in the PA OTOC is already an order-of-magnitude less than our target error of 0.02. In Fig.~\ref{fig:error}(b) we show how the error in the powder averaged OTOC compares to the error in the individual instances. Crucially, the error in the powder averaged OTOC is an order of magnitude less than the maximum error among the individual instances. In Fig.~\ref{fig:error}(c) we show the average error due to Haar-state sampling that is used to compute the trace averaged over 17 carbon-carbon pairs across two protein systems (TP901-1-DNA, TRalpha-RXR). The error is computed by comparing to the OTOC computed with $n_h = 160$ Haar samples. At $n_h = 20$, which is used throughout this paper, the error is already 2 orders of magnitude less than our target accuracy of 0.02. Note that our measurement of the sampling error may actually be limited by the single-precision simulation of the OTOC itself which is why it can't be converged below $\sim 5e-5$. Fig.~\ref{fig:error}(d) shows that the sampling error in the powder averaged OTOC is significantly lower than the max error among the individual instances as expected.

\subsection{More on Pauli propagation}\label{sec:App_pauli_extra}
\begin{figure}
    \centering
    \includegraphics[scale=0.7]{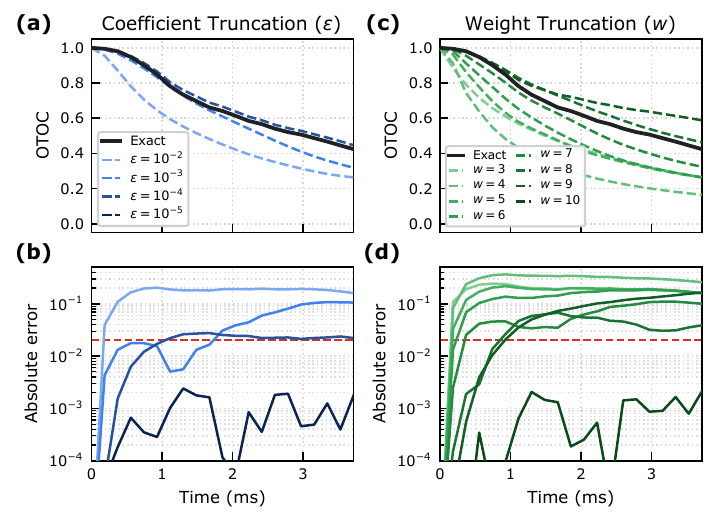}
    \caption{Approximate OTOCs and OTOC error from Pauli propagation based on (a),(b) coefficient truncation and (c),(d) weight truncation. Both converge to the exact result as expected, but the convergence with Pauli weight is not monotonic and full-weight Paulis have to be included to get results below an error threshold of 0.02. Results are shown for the Roquin-mRNA (NA-00) system for $n = 10$ spins.}
    \label{fig:pauli_extra}
\end{figure}

In the main text we show results for Pauli propagation with a fixed pool size (Fig.~\ref{fig:pauli_prop}), but many other truncation schemes are possible. In particular, Pauli propagation is most effective when truncation based on Pauli weight can be applied\cite{beguvsic2024fast,angrisani2025classically,rudolph2026pauli}. For the OTOC (see Eq.~\eqref{eqn:pauli_otoc}), there is no reason to think that the contribution from high-weight Paulis will be small, and this is what we observed in some small tests (shown in Fig.~\ref{fig:pauli_extra}). Truncation based on small coefficient values is very similar to the truncation to a fixed pool size which also truncates based on the size of the coefficients. In the main text, our analysis was based on the pool size because it is easier to analyze as a function of system size. In Fig.~\ref{fig:pauli_extra} we show a small example of these other truncation schemes.

\subsection{DTWA for OTOCs}\label{sec:App_dtwa}
Here we describe in more detail how we implement the DTWA to compute OTOCs. The method is the same as the method of Ref.~\cite{shirai2025out} with the notation of this section specialized to our use-case. Our goal is to compute the OTOC of the form
\begin{equation}
    \mathrm{OTOC} = \frac{1}{2^n}\mathrm{Tr}\left[X_b(t)X_mX_b(t)X_m \right],
\end{equation}
where $X$ is the Pauli $X$ operator. To evaluate this quantity using the DTWA, we follow the derivation of Ref.~\cite{shirai2025out} and first write the OTOC as
\begin{equation}
    \mathrm{OTOC} = \sum_{\gv{\tau}}W_{\gv{\tau}}\mathrm{Tr}\left[X_b(t)X_mX_b(t)X_m A(\mathbf{s}_{\gv{\tau}})\right].
\end{equation}
Note that
\begin{align}
    X_m A(\mathbf{s}_{\gv{\tau}}) &= \nonumber \\
    \frac{1}{2}[X_m &A(\mathbf{s}_{\gv{\tau}}) + A(\mathbf{s}_{\gv{\tau}})X_m+ X_m A(\mathbf{s}_{\gv{\tau}}) -  A(\mathbf{s}_{\gv{\tau}})X_m ]\nonumber \\
    &= \frac{1}{2}\{X_m,A(\mathbf{s}_{\gv{\tau}}) \} + \frac{1}{2}[X_m,A(\mathbf{s}_{\gv{\tau}})].
\end{align}
The advantage of this expression is that the first term is purely Hermitian while the second is purely anti-Hermitian so that we may split the OTOC into real and imaginary parts:
\begin{align}
    \mathrm{Re} &= \frac{1}{2}\sum_{\gv{\tau}}W_{\gv{\tau}}\mathrm{Tr}\left[X_b(t)X_mX_b(t)\{X_m, A(\mathbf{s}_{\gv{\tau}})\}\right] \\
    \mathrm{Im} &= -\frac{i}{2}\sum_{\gv{\tau}}W_{\gv{\tau}}\mathrm{Tr}\left[X_b(t)X_mX_b(t)[X_m, A(\mathbf{s}_{\gv{\tau}})]\right].
\end{align}
The imaginary part is zero at infinite temperature.
Despite the fact that the imaginary part is expected to be zero for the infinite-temperature OTOC considered here, it is still useful to compute it as a measure of error. With a little work one can show that
\begin{align}
    \{X_m, A(\mathbf{s}_{\gv{\tau}})\} &= 2s^x_{\tau_m}A(\mathbf{s}^1_{\gv{\tau}}) \\
    [X_m, A(\mathbf{s}_{\gv{\tau}})] &= 2i\left(A(\mathbf{s}^2_{\gv{\tau}}) - A(\mathbf{s}^3_{\gv{\tau}}) \right),
\end{align}
where we have defined specialized spin vectors:
\begin{align}
    (\mathbf{s}^1_{\gv{\tau}})_i &\equiv \begin{cases}
    \mathbf{s}_{\tau_i}, & i \neq m, \\
    s^x_{\tau_m}\hat{\mathbf{x}}, & i = m,
    \end{cases} \\
    (\mathbf{s}^2_{\gv{\tau}})_i &\equiv \begin{cases}
        \mathbf{s}_{\tau_i}, & i \neq m, \\
        \hat{\mathbf{x}}\times \mathbf{s}_{\tau_m}, & i = m,
    \end{cases} \\
    (\mathbf{s}^3_{\gv{\tau}})_i &\equiv \begin{cases}
        \mathbf{s}_{\tau_i}, & i \neq m, \\
        \mathbf{0}, & i = m.
    \end{cases}
\end{align}
Computation of the real and imaginary parts of the OTOC now comes down to three terms:
\begin{equation}
    G^p_{\gv{\tau}}(t) \equiv \mathrm{Tr}\left[X_b(t)X_mX_b(t)A(\mathbf{s}^p_{\gv{\tau}})\right], \quad p = 1, 2, 3.
\end{equation}
We will apply the DTWA to these objects. First, we rewrite it slightly as
\begin{align*}
    G^p_{\gv{\tau}}(t) &= \mathrm{Tr}\left[X_mX_b(t)A(\mathbf{s}^p_{\gv{\tau}})X_b(t)\right] \\
    &= \mathrm{Tr}\left[X_me^{iHt}X_be^{-iHt}A(\mathbf{s}^p_{\gv{\tau}})e^{iHt}X_b
    e^{-iHt}\right].
\end{align*}
Now, we can apply the DTWA,
\begin{equation}
    e^{-iHt}A(\mathbf{s}^p_{\gv{\tau}})e^{iHt} \approx A(\mathbf{s}^p_{\gv{\tau}}(t)),
\end{equation}
so that the phase-space variables are propagated forward in time to $t$. The next step is to apply the butterfly operators to yield
\begin{equation}
    X_b A(\mathbf{s}^p_{\gv{\tau}}(t))X_b \equiv A(\tilde{\mathbf{s}}^p_{\gv{\tau}}(t)),
\end{equation}
where we have used $\tilde{s}$ as a shorthand to indicate conjugation by $X_b$.
Finally, we apply the DTWA again but now backwards in time,
\begin{equation}
    e^{iHt}A(\tilde{\mathbf{s}}^p_{\gv{\tau}}(t))e^{-iHt} \approx A(\tilde{\mathbf{s}}^p_{\gv{\tau}}(0)).
\end{equation}
All that remains is to compute
\begin{equation}
    G^p_{\gv{\tau}}(t) = \mathrm{Tr}[X_m A(\tilde{\mathbf{s}}^p_{\gv{\tau}}(0))],
\end{equation}
and then plug this value into the expression for the real ($p = 1$) and imaginary ($p = 2, 3$) parts of the OTOCs.

\bibliography{ref}
\end{document}